\documentclass[11pt]{article} 
\usepackage{jcappub,natbib,float,caption,comment,bm}
\title{Constraints on primordial magnetic fields from the optical depth
of the cosmic microwave background}
\author[a]{Kerstin E. Kunze,}
\affiliation[a]{Departamento de F\'\i sica Fundamental and IUFFyM,
Universidad de Salamanca, Plaza de la Merced s/n, 37008 Salamanca,
Spain}
\author[b,c]{Eiichiro Komatsu}
\affiliation[b]{Max-Planck-Institut f\"ur Astrophysik,
 Karl-Schwarzschild-Str. 1, 85748 Garching, Germany}
\affiliation[c]{Kavli Institute for the Physics and Mathematics of the Universe (WPI), Todai Institutes for Advanced Study, The University of Tokyo,
 5-1-5 Kashiwanoha, Kashiwa, Chiba 277-8583, Japan}
\emailAdd{kkunze@usal.es}
\abstract{
Damping of magnetic fields via ambipolar diffusion and
decay of magnetohydrodynamical (MHD) turbulence in the post decoupling era
heats the intergalactic medium (IGM). Delayed recombination of hydrogen
atoms in the IGM yields an optical depth to scattering of the
cosmic microwave background (CMB). The optical depth generated at $z\gg 10$
does not affect the ``reionization bump'' of the CMB polarization power
spectrum at low multipoles, but affects the temperature and polarization
power spectra at high multipoles. Writing the present-day energy density
of fields smoothed over the damping scale at the decoupling epoch as
$\rho_{B,0}=B_{0}^2/2$, we constrain $B_0$ as a function of the spectral
index, $n_B$.
Using the Planck 2013 likelihood code that uses the Planck temperature and
lensing data together with the WMAP 9-year polarization data, we find the 95\%
upper bounds of $B_0<0.63$, 0.39, and 0.18~nG for $n_B=-2.9$, $-2.5$,
and $-1.5$, respectively. For these spectral indices, the optical depth
is dominated by dissipation of the decaying MHD turbulence that occurs
shortly after the decoupling epoch.  Our limits are 
stronger than the previous limits ignoring the
effects of the fields on ionization history. 
Inverse Compton scattering of CMB photons off electrons in the heated
IGM distorts the thermal spectrum of CMB. Our limits on $B_0$ imply that
the $y$-type distortion from dissipation of fields in the
post decoupling era should be smaller than $10^{-9}$, $4\times10^{-9}$, and $10^{-9}$, respectively.}
\begin{document}
\maketitle

\section{Introduction}
\setcounter{equation}{0}
Damping of magnetic fields affects the thermal and
ionization history of the universe. The damping processes before the
decoupling epoch can be approximately modeled by 
damping of  three magnetic modes. Of these,  damping of the fast
magnetosonic waves proceeds in a similar way as the Silk damping of
density perturbations, resulting in a damping scale on the order of the
Silk scale. Slow magnetosonic and Alfv\'en waves can avoid damping up to
smaller scales, resulting in a larger damping wave number. The value at
the decoupling epoch is given by \cite{jko1,sb}
\begin{eqnarray}
k_{d,dec}
=\frac{299.66}{\cos\theta}\left(\frac{B_0}{1~\rm nG}\right)^{-1}\;\; {\rm Mpc}^{-1},
\label{kdast}
\end{eqnarray}
for the  best-fit $\Lambda$CDM model of the ``Planck 2013+WP'' data \cite{Planck_cosmology}.
Here, $\theta$ is the angle between the wave vector and the
field direction which will be set to zero in this paper.

We assume that the magnetic field is a non-helical and Gaussian random 
field with its two-point function in Fourier space given by
\begin{eqnarray}
\langle B_i^*(\vec{k})B_j(\vec{q})\rangle=(2\pi)^3\delta({\vec{k}-\vec{q}})P_B(k)\left(\delta_{ij}-\frac{k_ik_j}{k^2}\right),
\label{2ptB}
\end{eqnarray}
where the power spectrum, $P_B(k)$, is assumed to be a power law,
$P_B(k)=A_Bk^{n_B}$, with the amplitude, $A_B$, and the spectral index,
$n_B$. 
This can be expressed in terms of the ensemble average of the 
present-day magnetic energy density by
\begin{eqnarray}
\langle\rho_{B,0}\rangle=\int\frac{d^3k}{(2\pi)^3}P_{B,0}(k)e^{-2\left(\frac{k}{k_c}\right)^2},
\end{eqnarray}
where $k_c$ is  a Gaussian smoothing scale. The power-law
magnetic power spectrum then gives
\begin{eqnarray}
P_{B,0}(k)=\frac{4\pi^2}{k_c^3}\frac{2^{(n_B+3)/2}}{\Gamma\left(\frac{n_B+3}{2}\right)}\left(\frac{k}{k_c}\right)^{n_B}\langle\rho_{B,0}\rangle.
\end{eqnarray}

For convenience, we define a smoothed magnetic field strength at the
present epoch, $B_{\lambda_c,0}$, as $\langle\rho_{B,0}\rangle\equiv
\frac{1}{2}B_{\lambda_c,0}^2$, where $\lambda_c\equiv 2\pi/k_c$ is a
wavenumber corresponding to $k_c$. In this paper, we
shall set $k_c$ to the maximal wave number determined by the 
damping scale at the decoupling epoch, $k_{d,dec}$, given by
equation~(\ref{kdast}). For simplicity we shall write $B_0\equiv
B_{\lambda_{d,dec},0}$ throughout the rest of the paper.

Primordial magnetic fields present before the decoupling epoch can
generate scalar, vector, and tensor modes of CMB temperature and polarization
anisotropies \cite{yikm,kb,pfp,sl,kk2}.
In general, there are two different types; namely, the compensated magnetic mode which is similar to an isocurvature
mode, and the passive mode due to the presence of anisotropic stress
of the magnetic field before neutrino decoupling. The latter is just like a curvature mode with its amplitude determined by the 
magnetic anisotropic stress and the time of generation of the primordial magnetic field. These can be important contributions to the CMB anisotropies, depending on the magnetic field
parameters. On small angular scales, the scalar compensated mode  
as well as the vector mode for nG fields with positive
spectral indices, $n_B>0$, 
can dominate over the primary CMB anisotropies. On large angular scales,
the tensor compensated mode for a few nG fields with a nearly scale invariant
spectrum, $n_B=-2.9$, can dominate \cite{Planck15_magnetic}. However, here we shall ignore these contributions and derive constraints
focusing only on the effect of dissipation of fields in the post
decoupling era. Since we only consider $n_B<0$, the former
contributions can be ignored. The limit we obtain for $n_B=-2.9$ is
smaller than nG, making the latter contributions sub-dominant. In any
case, including these contributions in the analysis would make the
constraints on $B_0$ stronger.

After the decoupling epoch, magnetic fields are damped by ambipolar
diffusion and decaying MHD turbulence \cite{sesu}. This leads to a change in the
ionization and thermal history, hence the visibility function of CMB
anisotropies. 
In ref.~\cite{kuko}, we have calculated the thermal and
ionization history of the IGM including dissipation of
fields. Ionization gives an optical depth to Thomson scattering of the
CMB. 
By writing the total optical depth as
$\tau_{tot}(B_0, n_B)=\tau_{tot}(B_0=0)+\Delta\tau(B_0,n_B)$, 
the additional contribution from dissipation of magnetic fields 
$\Delta\tau$ is given by\footnote{This formula is a minor update of the formula we gave in
ref.~\cite{kuko}. We now use the Planck 13+WP best fit cosmological
parameters. Following ref.~\cite{Planck15_magnetic,Chluba:2015lpa}, we
calculate the photoionization rates at the photon
temperature instead of the electron temperature in the modified version
of {\tt Recfast++},  and include the collisional
ionization.  Finally, we have also corrected a numerical error in the amplitude 
of the ambipolar diffusion heating rate. We are grateful to
J. Chluba for helping us implement these updates.}
\begin{eqnarray}
\Delta\tau(B_0, n_B)&=&0.0241\left(\frac{B_0}{\rm nG}\right)^{1.547}(-n_B)^{-0.0370}
\nonumber\\
&\times&{\rm e}^{
-5.2815\times 10^{-12}(-n_B)^{23.8731}+5.4\times 10^{-3}\left(\frac{B_0}{\rm nG}\right)^{3.3706}
-7.1\times 10^{-3}(-n_B)^{1.948}
\left(\frac{B_0}{\rm nG}\right)^{2.0713}}
\end{eqnarray}
for $n_B<0$.

In general, there are two important contributions to the total optical
depth out to the decoupling epoch: the contribution due to reionization,
$\tau_{reio}$, and that generated at high redshifts close to decoupling,
$\tau_{high-z}$. Hence $\tau_{tot}=\tau_{reio}+\tau_{high-z}$.
Whereas $\tau_{reio}$ determines the CMB polarization angular power spectrum at 
low multipoles  (``reionization bump''), the total optical depth out to decoupling determines 
the overall suppression of the temperature power spectrum at  
high multipoles as $C_\ell^{TT}\to C_\ell^{TT}e^{-2\tau_{tot}}$.
As the contribution to the
optical depth from dissipation of fields comes from high redshifts,
$z\gg 10$, it has little effect on $\tau_{reio}$ and hence
the CMB polarization power spectrum at low multipoles. Therefore, the change in the amplitude of 
the temperature angular power spectrum induced by the magnetic field dissipation is determined
by $\Delta\tau(B_0,n_B)$, and the temperature data at high multipoles
will be the key to measure this effect. 
In this paper, we shall use
the Planck 2013 temperature data
\cite{Planck_overview,Planck_likelihood} (including the lensing data
\cite{Planck_lens}) and the Planck 13 polarization  likelihood \cite{Planck_likelihood}
derived from the WMAP 9-year polarization data
\cite{bennett/etal:2013,wmap9} to constrain $B_0$ for some
representative values of $n_B$. In particular, the lensing data play an
important role in isolating the optical depth in the temperature data.

The rest of the paper is organized as follows.
In section \ref{sec_CMB}, we calculate the CMB temperature and
polarization power spectra with a modified thermal and reionization
history due to the dissipation of the magnetic field in the post
decoupling era. In section \ref{sec_par-est}, we obtain constraints on
the magnetic field strength for some representative values
of $n_B$. We conclude in section \ref{sec_conc}.

\section{CMB anisotropies with dissipation of magnetic fields in the
 post-decoupling era}
\setcounter{equation}{0}
\label{sec_CMB}

Dissipation of magnetic fields in the post decoupling era takes place by
two processes; namely, ambipolar diffusion and decaying MHD turbulence
\cite{sesu}. These processes affect the evolution of the electron
temperature, $T_e$, as
\begin{eqnarray}
\dot{T}_e=-2\frac{\dot{a}}{a}T_e+\frac{x_e}{1+x_e}\frac{8\rho_{\gamma}\sigma_T}{3m_ec}\left(T_{\gamma}-T_e\right)+\frac{x_e\Gamma}{1.5 k_B n_e},
\end{eqnarray}
where the dissipation rate, $\Gamma=\Gamma_{\rm in}+\Gamma_{\rm decay}$,
includes the contribution from  ambipolar diffusion, $\Gamma_{\rm in}$,
and decaying MHD turbulence, $\Gamma_{\rm decay}$. The other
variables are: $a$ the scale factor, $T_\gamma$ the photon temperature,
$\sigma_T$ the Thomson scattering cross section, $x_e$ the ionization
fraction, $n_e$ the electron number density, $\rho_\gamma$ the photon
energy density, $m_e$ the electron rest mass, $c$ the speed of light,
and $k_B$ the Boltzmann constant.

Ambipolar diffusion is due to different velocities of the neutral and
the ionized matter components in the presence of a magnetic field. As
this velocity difference is caused by the action of the Lorentz force on
the ionized component, the dissipation rate of ambipolar diffusion is
determined by the Lorentz term as \cite{sesu}
\begin{eqnarray}
\Gamma_{\rm
 in}=\frac{\rho_n}{16\pi^2\gamma\rho_b^2\rho_i}|(\vec{\nabla}\times\vec{B})\times\vec{B}|^2,
\label{eq:gammain}
\end{eqnarray}
where $\rho_n$, $\rho_i$, and $\rho_b$ are the energy densities of
neutral hydrogen, ionized hydrogen, and the total baryons,
respectively, and $\gamma$ is  the coupling between the ionized and neutral
component given by $\gamma\simeq\langle\sigma v\rangle_{H^+,H}/(2m_H)$ with $\langle\sigma v\rangle_{H^+,H}=0.649~T^{0.375}\times
10^{-9}$~cm$^3$~s$^{-1}$ \cite{sbk2}. 
Equation (\ref{eq:gammain}) is evaluated for the average Lorentz force using the two point function of the magnetic field (cf., eq. (\ref{2ptB})).
The final expression depends on the magnetic field strength as well as
on its spectral index, and can be found in \cite{kuko}. 

Dissipation of the magnetic field due to decaying MHD turbulence rests
on the fact that  turbulence is no longer suppressed in
the plasma after the decoupling epoch. On scales below the magnetic
Jeans scale,  the magnetic energy on large scales is transferred to small scales and dissipates. The  dissipation
rate of this highly non linear process is estimated by numerical
simulations of MHD turbulence in flat space using the fact that there
exists an appropriate rescaling of variables in flat space to match
those in an expanding, flat
Friedmann-Robertson-Walker background \cite{decMHDsim1,decMHDsim2,decMHDsim3,decMHDsim4,decMHDsim5,decMHDsim6}.
In the matter-dominated era, the
estimated dissipation rate of a non-helical magnetic field is given by
\cite{sesu}
\begin{eqnarray}
\Gamma_{\rm decay}=\frac{B_0^2}{8\pi}\frac{3m}{2}\frac{\left[\ln\left(1+\frac{t_d}{t_i}\right)\right]^m}
{\left[\ln\left(1+\frac{t_d}{t_i}\right)+\ln\left[\left(\frac{1+z_i}{1+z}\right)^{\frac{3}{2}}\right]\right]^{m+1}}
H(t)(1+z)^4,
\label{eq:gammadecay}
\end{eqnarray}
where $B_0$ is the present-day field value assuming a flux
freezing; $m$ is related to the magnetic spectral index as
$m=\frac{2(n_B+3)}{n_B+5}$; $t_d$ is the physical decay time scale for
turbulence given by $t_d/t_i=(k_J/k_d)^{\frac{(n_B+5)}{2}}
\simeq 14.8~(B_0/1~{\rm nG})^{-1}(k_d/1~{\rm
Mpc}^{-1})^{-1}$ 
with the magnetic Jeans wavenumber of $k_J\simeq
14.8^{\frac{2}{n_B+5}} (B_{0}/{\rm
1~nG})^{-\frac{2}{n_B+5}}\left(k_d/1~{\rm
Mpc}^{-1}\right)^{\frac{n_B+3}{n_B+5}}$ Mpc$^{-1}$ \cite{sesu}; and
$z_i<z_{dec}$ and $t_i>t_{dec}$ are the 
redshift and time at which dissipation of the magnetic field due to
decaying MHD turbulence becomes important.

Ref.~\cite{sesu} calculated the evolution of the matter temperature,
$T_e$, and the ionization fraction, $x_e$, with the effect of damping of
magnetic fields in the post decoupling era. In
ref.~\cite{kuko}, we revisited these calculations using a
modified version of {\tt Recfast++}
\cite{recfast1,recfast2,recfast3,recfast4,recfast5,recfast6,recfast7}. 
We found that ambipolar diffusion contributes at 
$z\lesssim 100$, and the effect decreases as $n_B$ decreases. The amplitude is
proportional to $B_0^4$. On the
other hand, the effect of decaying MHD turbulence increases as $n_B$
decreases, is proportional to $B_0^2$, and dominates over ambipolar
diffusion up to intermediate redshifts.

\begin{figure}[t]
\centerline{\epsfxsize=2.9in\epsfbox{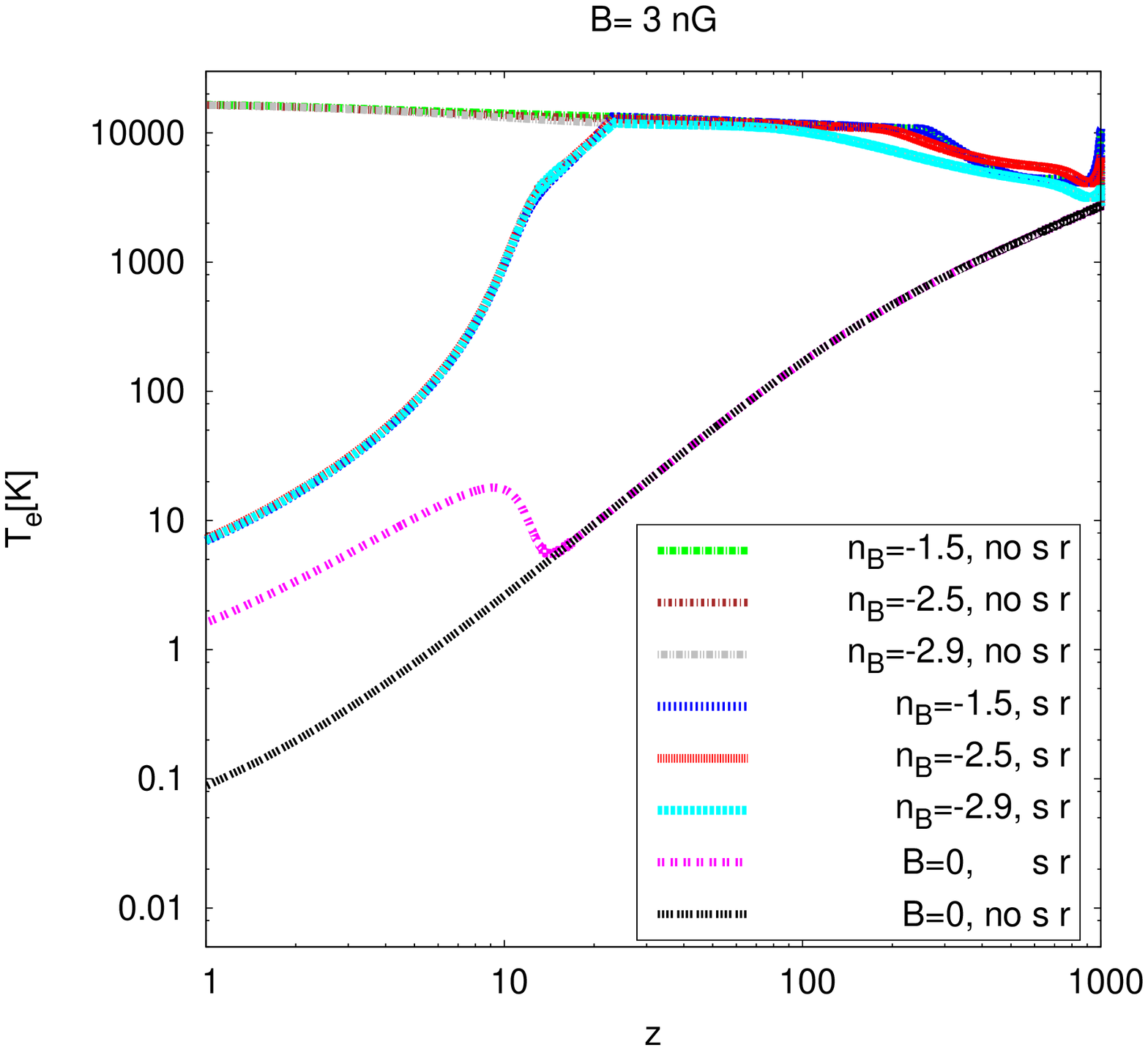}
\hspace{0.9cm}
\epsfxsize=2.9in\epsfbox{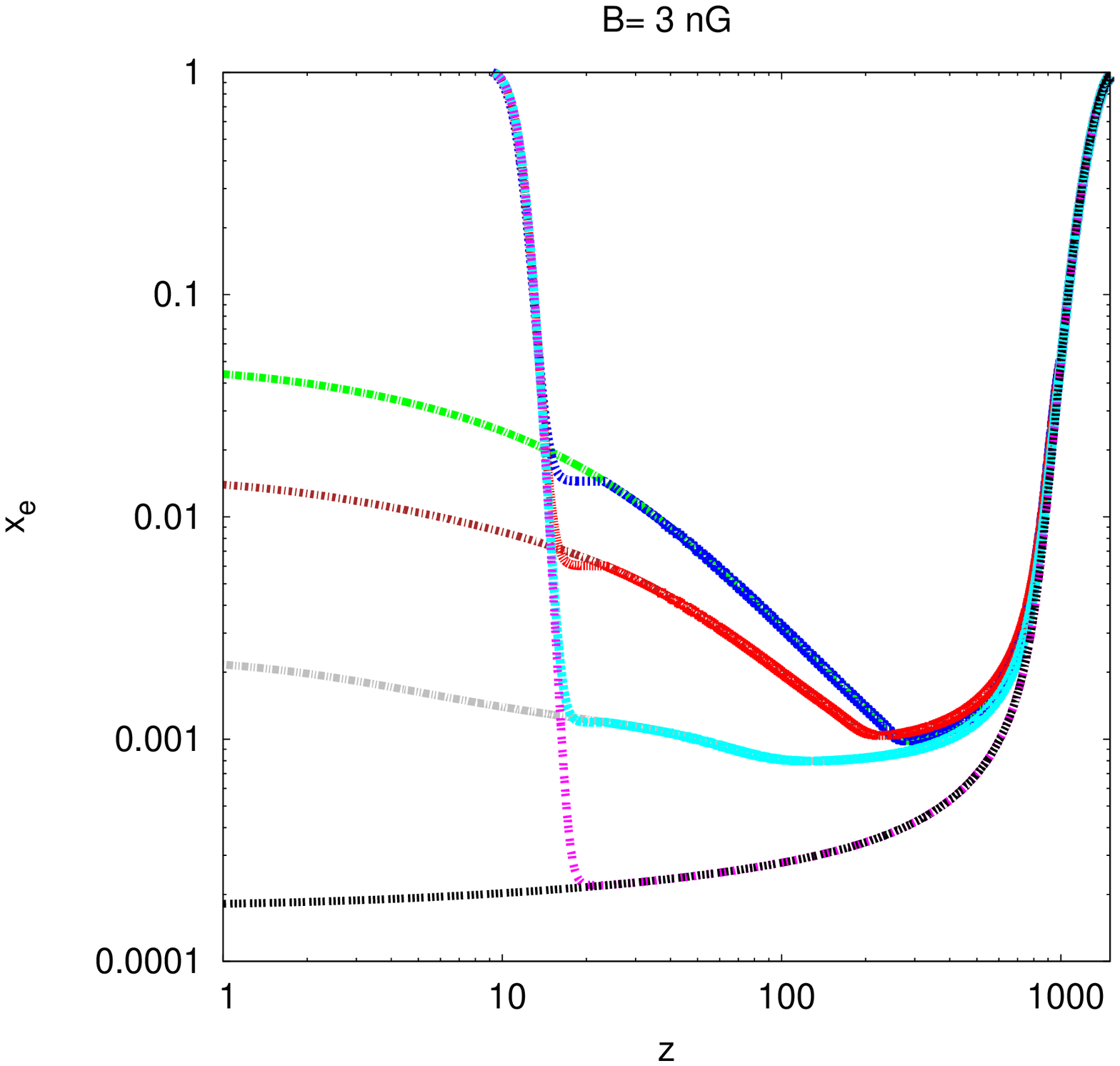}
}
\centerline{\epsfxsize=2.9in\epsfbox{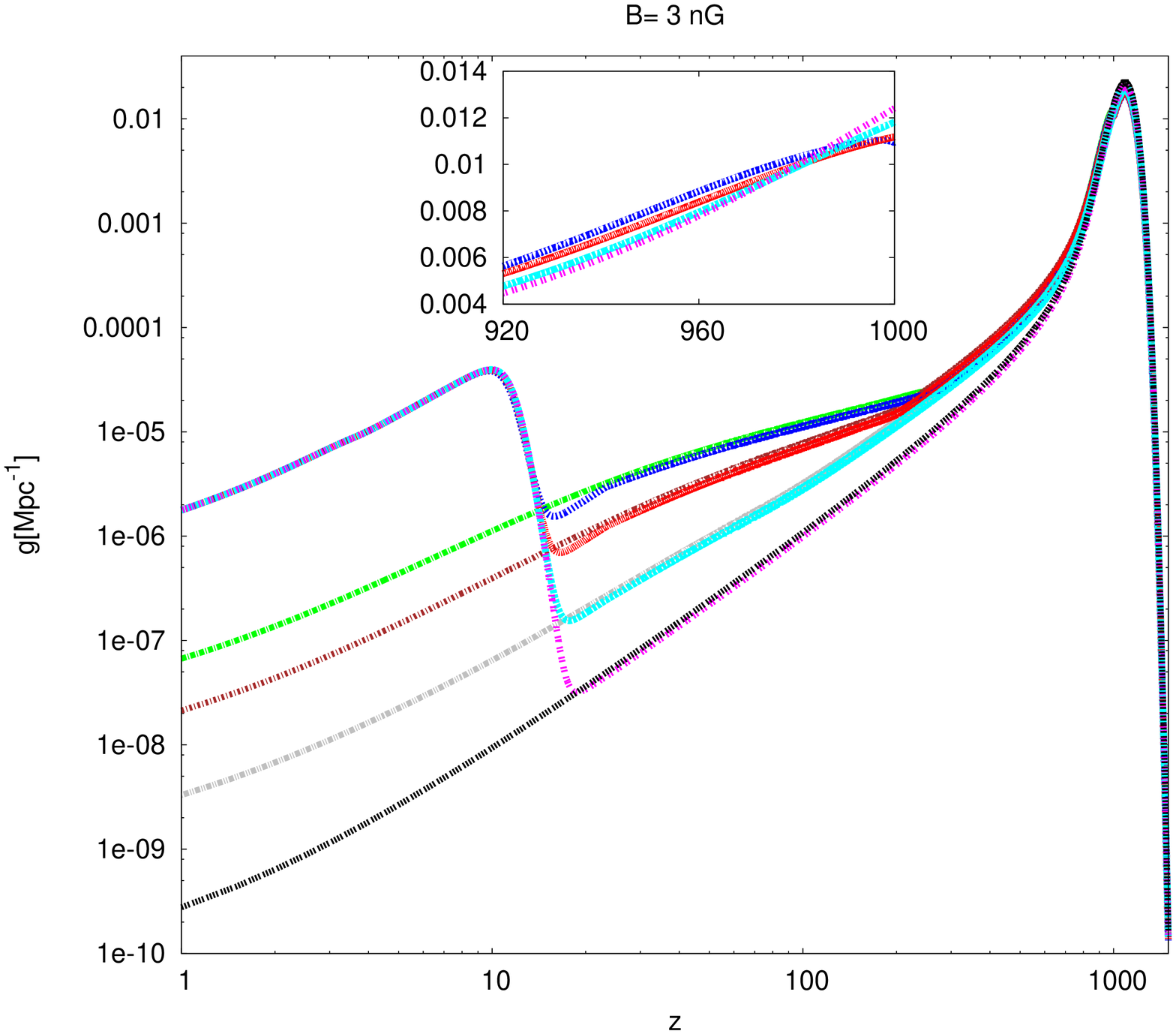}
\hspace{0.9cm}
\epsfxsize=2.9in\epsfbox{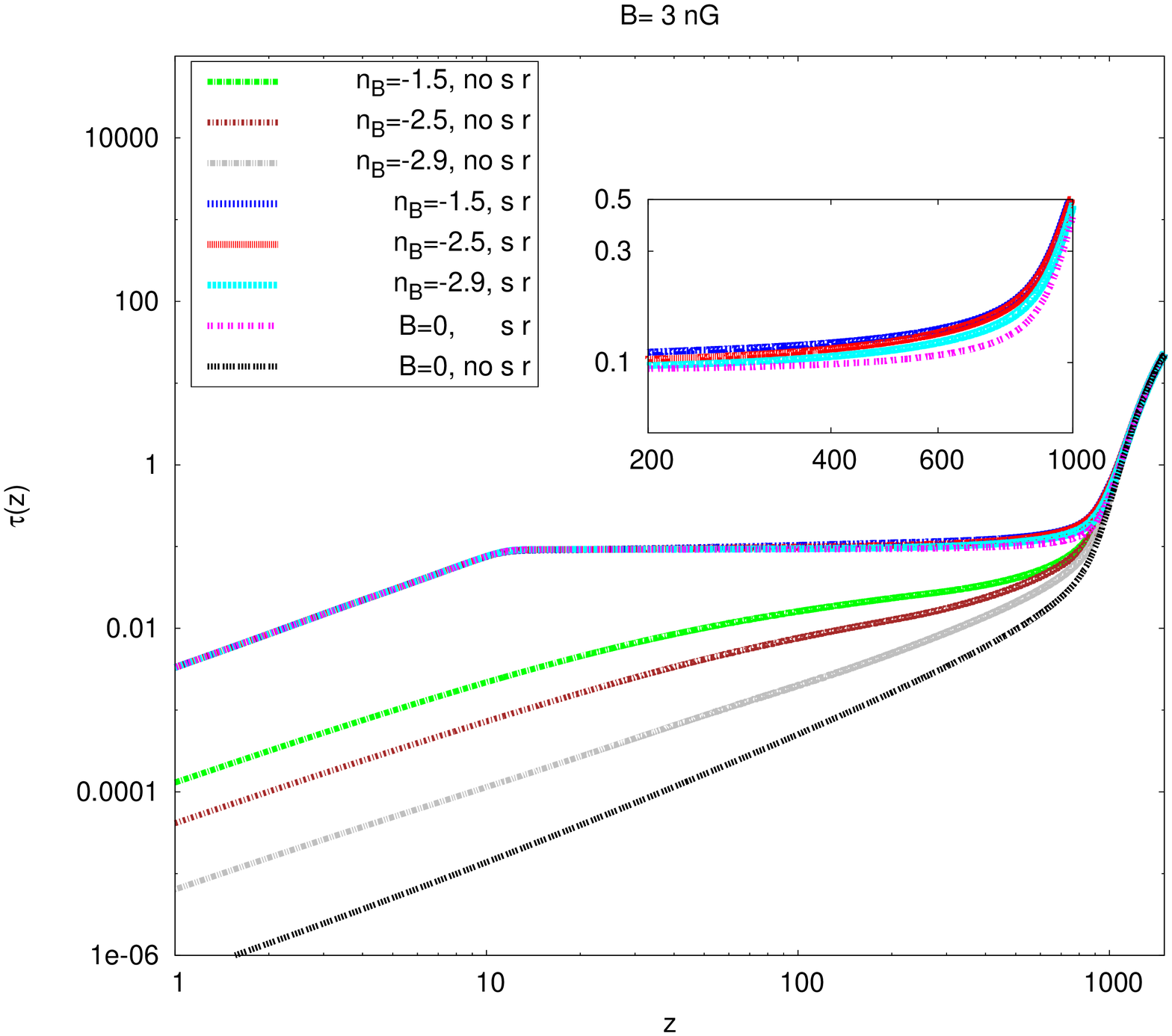}
}
\caption{Evolution of the matter temperature ({\it top left}), the ionization
 fraction ({\it top right}), the visibility function of CMB ({\it bottom left}), and
 the total optical depth to scattering of  CMB ({\it bottom right}), with and
 without dissipation of magnetic fields, and with and without the
 standard late-time reionization (indicated by ``s r'' in the
 legend) of the universe with 
 $\tau_{reio}=0.0925$. For better clarity the legend has been included only in two 
 of the figures, though it applies to all four figures.
 To see the effect of dissipation of
 fields in the standard late-time reionization model, compare the blue, red, cyan
 and pink lines. The cosmological parameters are the best fit values of 
 Planck 2013 data plus the low-$\ell$ polarization data of WMAP 9
 \cite{Planck_cosmology}. The field strength is $B_0=3 $~nG
 with spectral indices of $n_B= -1.5$, $-2.5$, and $-2.9$. 
 At $z\gtrsim 20$, dissipation
 is dominated by the decaying MHD turbulence.
 The inset in the bottom-right panel shows an increase in the
 total optical depth due to decaying MHD turbulence at $z>200$.
 }
\label{fig1}
\end{figure}

In figure \ref{fig1} we show the evolution of $T_e$, $x_e$, the
visibility function of CMB ($g$), and the optical depth to scattering of 
CMB ($\tau$), as a function of redshift. 
For each choice of parameters, we show the case including
the standard implementation of  instantaneous reionization (denoted as
``s r'') and not including it (denoted as ``no s r'').
We use the {\tt CLASS} code \cite{class1,class2,class3,class4} to
compute these quantities. The standard implementation of {\tt CLASS} is
the same as in the {\tt CAMB} code \cite{camb,Lewis}, which uses a tanh fitting formula for the ionization fraction centered at the 
reionization redshift.
The case not including the standard reionization is shown only for
comparison.
Dissipation of
the decaying turbulence heats the IGM since shortly after the decoupling
epoch, delaying recombination of hydrogen atoms in the IGM. At low
redshifts, $z\lesssim 20$, the temperature of the completely reionized universe
falls as $z$ decreases because ambipolar diffusion cannot act on ionized
plasma. In a complete treatment of the late time thermal and ionization history
one should also include contributions from stellar feedback which
ionizes and can efficiently heat the IGM up to temperatures of  $10^4$~K
at $z\lesssim 10$. 
Since our main conclusions are not based on the
temperature but on the ionization fraction, the error in the temperature
evolution at $z\lesssim 20$ has no consequences. Decay of MHD turbulence
increases the total optical depth to scattering of CMB shortly after the
decoupling epoch, and this changes the power spectra of the temperature
and polarization anisotropies, as discussed below.

 \begin{figure}[t]
\centerline{\epsfxsize=3.3in\epsfbox{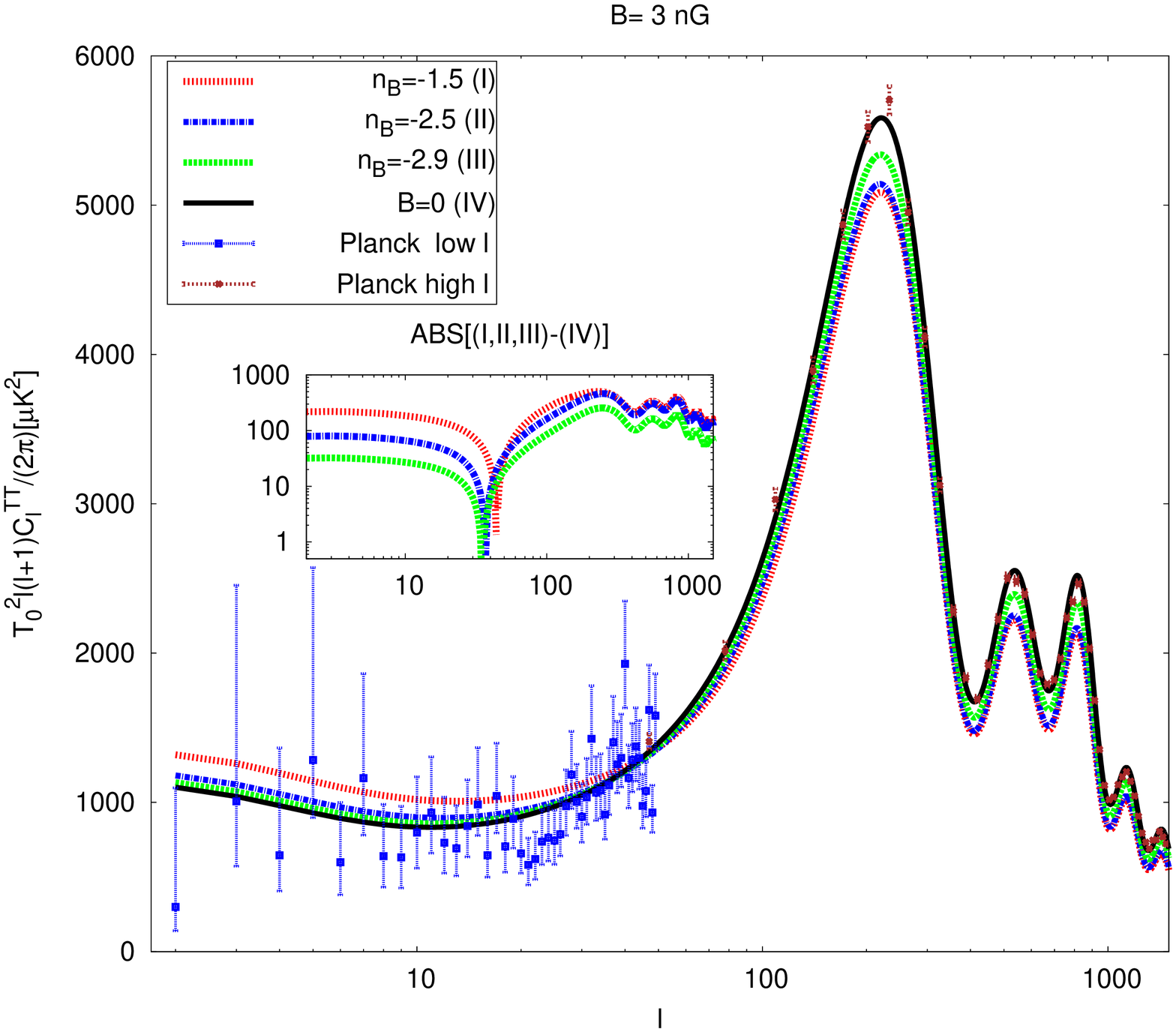}
\hspace{0.1cm}
\epsfxsize=3.3in\epsfbox{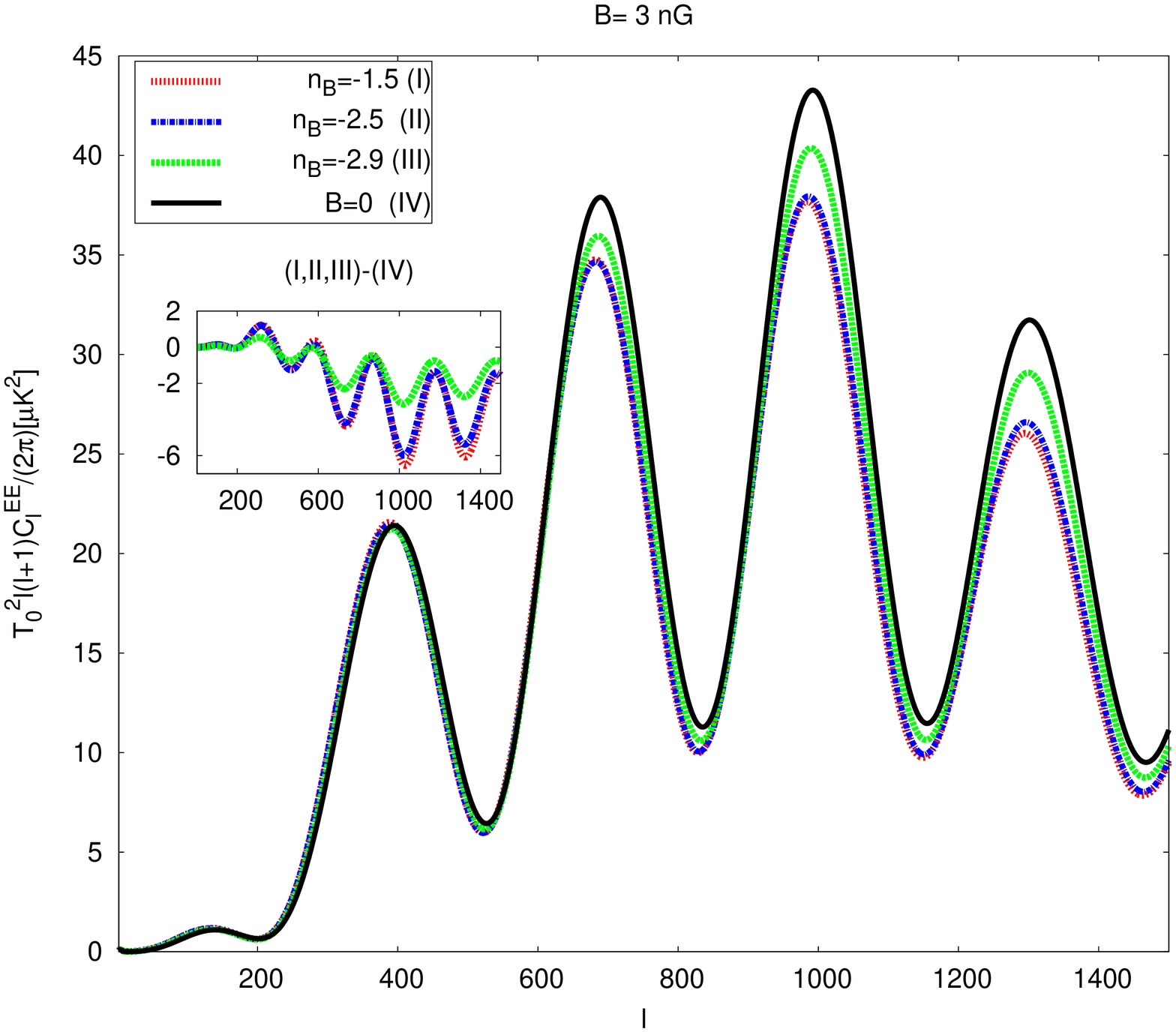}}
\centerline{\epsfxsize=3.3in\epsfbox{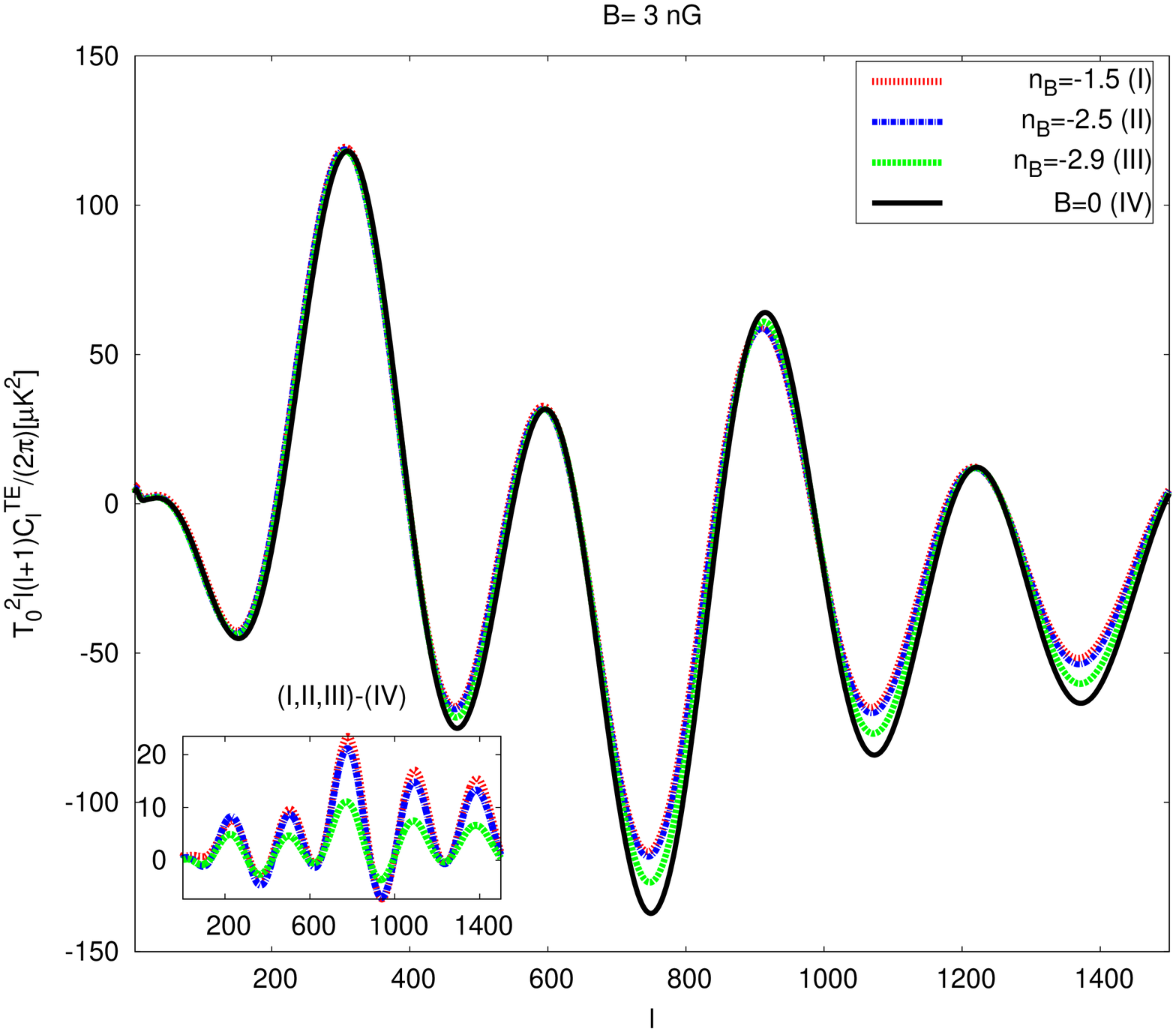}}
\caption{Temperature and polarization power spectra of CMB, with and
  without dissipation of magnetic fields, assuming  standard late-time reionization of the universe with
  $\tau_{reio}=0.0925$. The magnetic field parameters are $B_0=3$~nG
  for three different values of its spectral index  $n_B$. 
  The insets show the difference between each
  choice of spectral index and the case without the magnetic field.
    The additional optical depth to scattering
  of CMB due to dissipation of the fields suppresses the power spectra
  at high multipoles, while regenerating the polarization power spectra
  in $100\lesssim \ell\lesssim 400$. Also, the shift of the visibility
  function to lower redshifts shifts the locations of the acoustic peaks
  of the power spectra to lower multipoles. 
The cosmological parameters are the best fit values of 
 Planck 2013 data plus the Planck 13 likelihood of the low-$\ell$ polarization data of WMAP 9 \cite{Planck_cosmology}.
 We also show the Planck 2013 temperature data \cite{Planck_overview}.}
\label{fig2}
\end{figure}

We use the {\tt CLASS} code with its standard implementation of instantaneous reionization at late times to
calculate the power spectra, and show the effects of dissipation of
fields in figure \ref{fig2} for a magnetic field with$B_0= 3$~nG
 and
spectral indices of $n_B= -2.9$, $-2.5$, and $-1.5$.
It is well known that the optical depth to
scattering of CMB suppresses the temperature and polarization power
spectra by $e^{-2\tau_{tot}}$ at multipoles higher than that
corresponding to the horizon size at the epoch of the scattering. 
In addition to the suppression due to the standard late-time
reionization, the  optical
depth due to dissipation of fields, $\Delta\tau$,  creates an additional
suppression of the power spectra by $e^{-2\Delta\tau}$ (see figure~\ref{fig2}).
 Also, as the peak of
the visibility function of CMB shifts slightly toward a lower redshift
than the standard decoupling redshift, the angular diameter distance to
the decoupling epoch is slightly reduced, shifting the locations of the
acoustic peaks of the temperature and polarization power spectra to
lower multipoles. 
 This effect is more pronounced for polarization, and is clearly
visible in the top right panel of figure \ref{fig2}.

Not only does an additional optical depth suppress the power spectra at
multipoles higher than that corresponding to the horizon size at the
epoch of the scattering, but also creates an additional polarization
{\it at} multipoles corresponding to the horizon size at the
epoch of the scattering. The amplitudes of the first and second peaks in
the $E$-mode polarization power spectrum at $\ell\approx 100$ and $400$
are enhanced due to regeneration of polarization by
additional scattering of CMB between the decoupling  and the
reionization epochs. This can be seen in the top right panel of figure 
\ref{fig2}. While  the polarization power spectrum at higher multipoles 
is significantly influenced  by the change in the optical depth,
that at lower multipoles is not. 
The  ``reionization bump'' in the polarization power
 spectrum at $\ell\lesssim 10$ is not affected by dissipation of the
 fields, as ionization at $z\lesssim 20$ is totally dominated by that of
 the late-time  reionization.
 Similar features are observed in the
temperature-polarization cross power spectrum 
(cf. lower panel of figure \ref{fig2}).
This phenomenology is similar to the effect of heating due to
annihilation of dark matter particles \cite{decDM0,decDM1, decDM2,
Planck_cosmology}.

\begin{figure}[t]
\centerline{\epsfxsize=3.3in\epsfbox{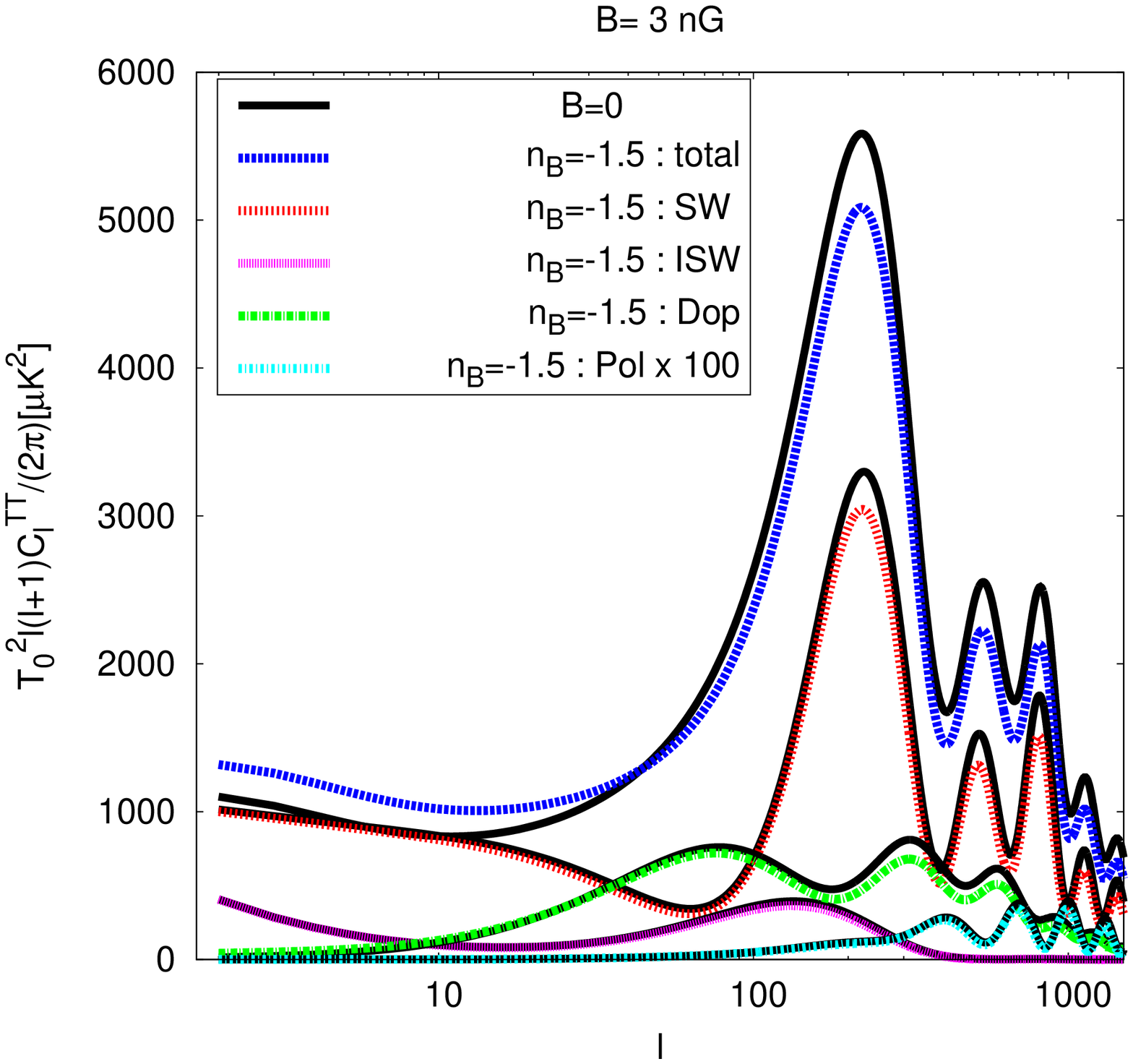}
\hspace{0.1cm}
\epsfxsize=3.3in\epsfbox{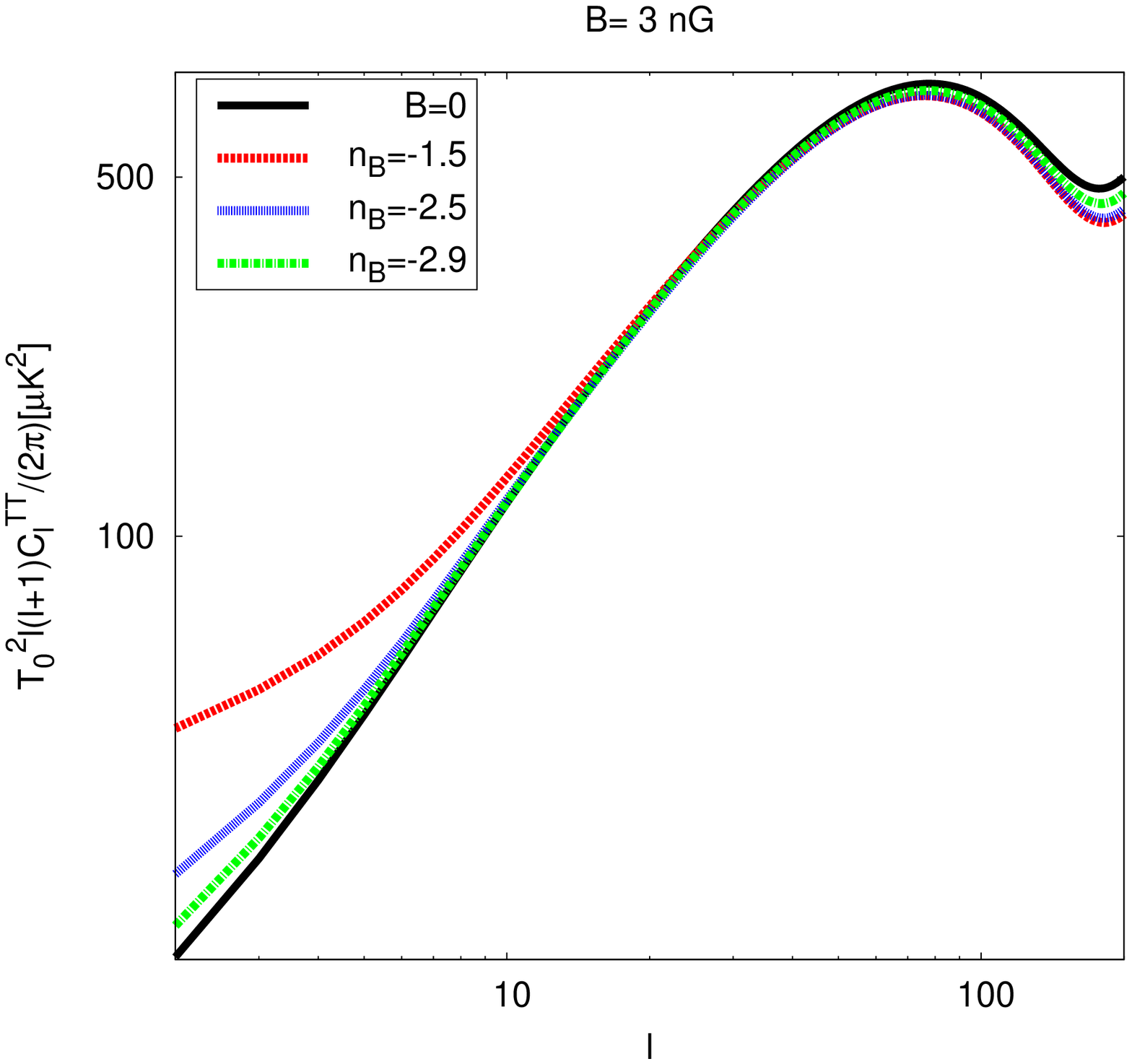}}
\caption{
Temperature power spectra of CMB, with and
  without dissipation of magnetic fields, assuming the standard late-time reionization of the universe with
  $\tau_{reio}=0.0925$. ({\sl Left}) Contributions from the Sachs-Wolfe term (SW), the total
  integrated Sachs-Wolfe effect (ISW) including early and late time
 contributions, the Doppler term (Dop) and the polarization term (Pol). 
   The black lines show the cases with no field, whereas the other lines
 show $B_0= 3$ nG and $n_B=-1.5$.
 ({\sl Right)} The Doppler term for $B=3 $nG and $n_B=-1.5$, $-2.5$, and
 $-2.9$.}
\label{fig2-add}
\end{figure}

The top left panel of figure \ref{fig2} shows that the low-$\ell$
temperature power spectrum increases with increasing magnetic spectral index.
We find that this
is due to an enhanced Doppler term (in Newtonian gauge). 
For our choice of parameters this is most noticeable for the largest
spectral index, $n_B=-1.5$.
In the left panel of figure~\ref{fig2-add}, we show the contributions from the Sachs-Wolfe term,
the integrated Sachs-Wolfe term, the Doppler term, and the polarization
term to the total angular power spectrum. 
The right panel shows only the Doppler term, which increases as $n_B$
increases at $\ell\lesssim 10$. This explains what we find in the top
left panel of figure~\ref{fig2}.

\section{Parameter estimation}
\setcounter{equation}{0}
\label{sec_par-est}

To estimate the magnetic field parameters along with
the cosmological parameters, we use Markov chains Monte Carlo as
implemented in {\tt montepython} \cite{montepython}. 
Specifically, we fix the magnetic
spectral index, $n_B$, and vary the field strength, $B_0$, along with the
six standard $\Lambda$CDM parameters, i.e., the Hubble constant,
$H_0=100~h~{\rm km/s/Mpc}$;
the physical baryon density, $\omega_{b}\equiv \Omega_bh^2$; the physical
cold dark matter density, $\omega_{cdm}\equiv \Omega_{cdm}h^2$; the
amplitude of the scalar power spectrum at $k=0.05~{\rm
Mpc}^{-1}$, $A_s$; the tilt of the power spectrum, $n_s$; and the optical depth
from the late-time reionization, $\tau_{reio}$.

We use the Planck 2013 temperature data
\cite{Planck_overview,Planck_likelihood} (including the lensing data
\cite{Planck_lens}) and the 
likelihood of the
WMAP 9-year polarization data
\cite{bennett/etal:2013,wmap9}
derived by the Planck collaboration \cite{Planck_likelihood}.
The lensing data are essential for breaking degeneracy between $A_s$ and $B_0$. This
degeneracy arises because the optical depth constraint from the
temperature data is degenerate with $A_s$. (The temperature data
effectively measure $A_se^{-2\tau_{tot}}$.)
In our study the key element is that the  magnetic field dissipation
significantly changes the optical depth close to decoupling and so changes the total 
optical depth to decoupling but does not change the optical depth to reionization.
 The former is determined by the temperature anisotropy data and the latter by the 
 polarization data.  The combination of the two data sets together with the lensing data
 allows to constrain the magnetic field parameters.

\begin{figure}[t]
\centerline{\epsfxsize=1.6in\epsfbox{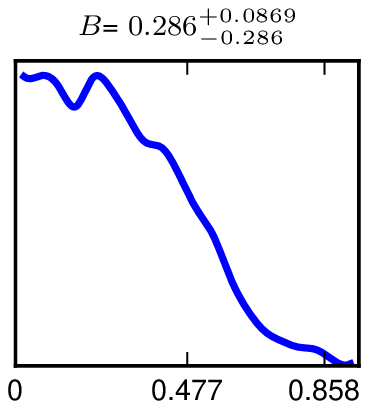}
\hspace{1.3cm}
\epsfxsize=1.6in\epsfbox{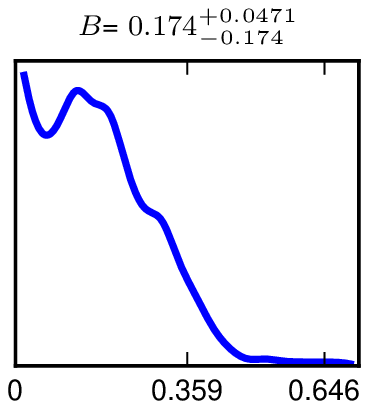}
\hspace{1.3cm}
\epsfxsize=1.6in\epsfbox{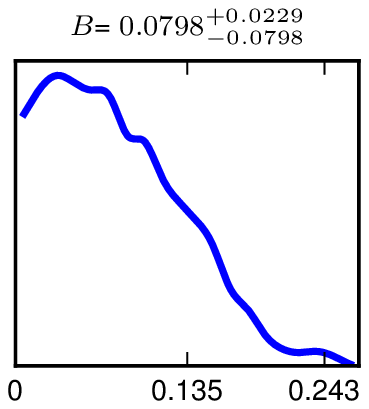}
}
\caption{Marginalized posterior distributions of the magnetic field
 strength, $B_0$ (in units of nG), for the magnetic
spectral index of $n_B=-2.9$ ({\it left}), $n_B=-2.5$ ({\it middle}) and $n_B=-1.5$ ({\it right}). 
}
\label{fig4}
\end{figure}

We summarize the constraints on $B_0$ and the cosmological parameters in
table \ref{tab1}, and show the marginalized posterior distributions of
$B_0$ in figure \ref{fig4}. 
We chose $n_B=-2.9$, $-2.5$, and $-1.5$ as representative cases.
We find no evidence for dissipation of the magnetic fields
in the data. The 95\% CL upper bounds on the field strength are
$B_0<0.63$, 0.39, and 0.18~nG for $n_B=-2.9$, $-2.5$, and $-1.5$,
respectively.\footnote{The corresponding smoothing scales are much
smaller than 1~Mpc. The largest one is of order 13 kpc for $B_0=0.63$~nG, and 4~kpc for $B_0=0.18$ nG (cf. equation (\ref{kdast})).}

These limits are significantly  stronger than those from
the magnetic scalar- and vector-mode contributions to the angular power
spectra of the CMB.  For example, the Planck collaboration
used the Planck 2013 temperature data, the WMAP 9-year polarization
data, and the high-$\ell$ temperature data of ACT and SPT to 
find $B_{1{\rm Mpc}}< 3.4$~nG (95\%~CL) for a smoothing scale of 1 Mpc
from  the magnetic scalar and vector modes, while
ignoring the effects on the ionization history of the universe.
Our limits on $B_0$, which is smoothed over the magnetic damping scale
at decoupling, 
become stronger if we scale them to a Mpc scale, as
changing the smoothing scale, $\lambda_s$, to a different smoothing
scale, $\lambda_*$, leads to a rescaling of the magnetic field amplitude
by a factor of $(\lambda_s/\lambda_*)^{\frac{n_B+3}{2}}$.

The only limit in the literature that is stronger than ours 
is $B_0<0.01$~nG for a smoothing scale of 0.1 kpc \cite{ja}, which was
derived from the change in the ionization history {\it before} the
decoupling epoch, due to the effect of clumping in the baryon density
perturbations induced by a primordial magnetic field. Rescaling our
limits to 0.1~kpc makes their limit much stronger than ours.

\begin{table}[t]
\begin{tabular}{|l|cc|cc|cc|} 
 \hline 
                                     & \hspace{0.5cm}$n_B=-2.9$         &                &\hspace{0.5cm}  $n_B=-2.5$         &                    &\hspace{0.5cm} $n_B=-1.5$& \\ \hline 
                                     & best-fit & 68\% limits & best-fit & 68\% limits   & best-fit & 68\% limits \\ \hline
$B_0$ & 0.2176&  $0.286_{-0.29}^{+0.087}$                                      &0.154&$0.1735_{-0.17}^{+0.047}$                              & 0.06024&$0.07979_{-0.08}^{+0.023}$\\ 
$100~\omega_{b }$ &2.208 &$2.214_{-0.029}^{+0.027}$                   & 2.21& $2.215_{-0.029}^{+0.027}$                                & 2.229& $2.214_{-0.028}^{+0.028}$ \\ 
$\omega_{cdm }$ &  0.12& $0.1188_{-0.0022}^{+0.0022}$             & 0.119& $0.1188_{-0.0022}^{+0.0022}$                         &  0.1183& $0.1188_{-0.0022}^{+0.0022}$\\ 
$H_0$ &  67.22& $67.78_{-1.1}^{+1}$                                                   &67.66& $67.78_{-1.1}^{+1}$                                           & 68.14& $67.74_{-1}^{+1}$\\ 
$10^{9}A_{s }$ & 2.172& $2.194_{-0.053}^{+0.05}$                              &2.225& $2.196_{-0.056}^{+0.048}$                                 &2.188& $2.199_{-0.056}^{+0.048}$ \\ 
$n_{s }$ &  0.9604& $0.9628_{-0.0069}^{+0.007}$                           &  0.9663& $0.9626_{-0.007}^{+0.0067}$                        &0.9635&  $0.9639_{-0.0075}^{+0.0074}$ \\ 
$\tau_{reio }$ &  0.08346& $0.08966_{-0.014}^{+0.012}$                   & 0.09624& $0.09001_{-0.014}^{+0.012}$                         &0.08873&$0.08969_{-0.014}^{+0.012}$ \\ 
\hline 
$-\ln{\cal L}_\mathrm{min}$& &4906.72& & 4906.63&& 4906.72 \\
$\chi^2_\mathrm{min}$ && 9813 && 9813 && 9813\\
\hline
 \end{tabular} \\ 
\caption{Best-fit values and 68\% confidence limits  on the present-day
 magnetic field strength, $B_0$ (in units of nG), smoothed over
 $k_{d,dec}$ given in  equation~(\ref{kdast}), and the standard
 $\Lambda$CDM cosmological  parameters.}
\label{tab1}
\end{table}

\begin{figure}[t]
\centerline{\epsfxsize=1.15in\epsfbox{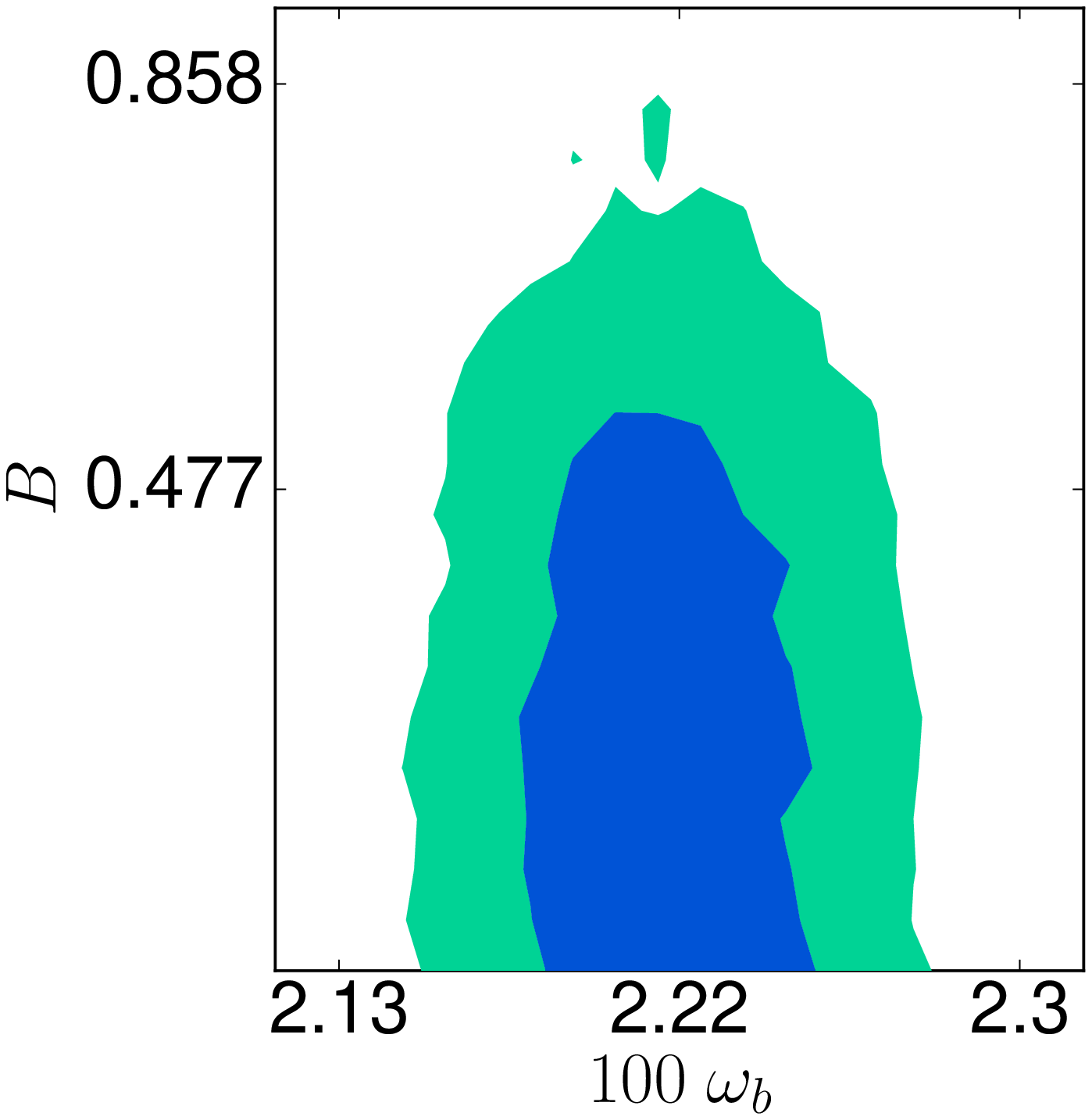}
\epsfxsize=1.15in\epsfbox{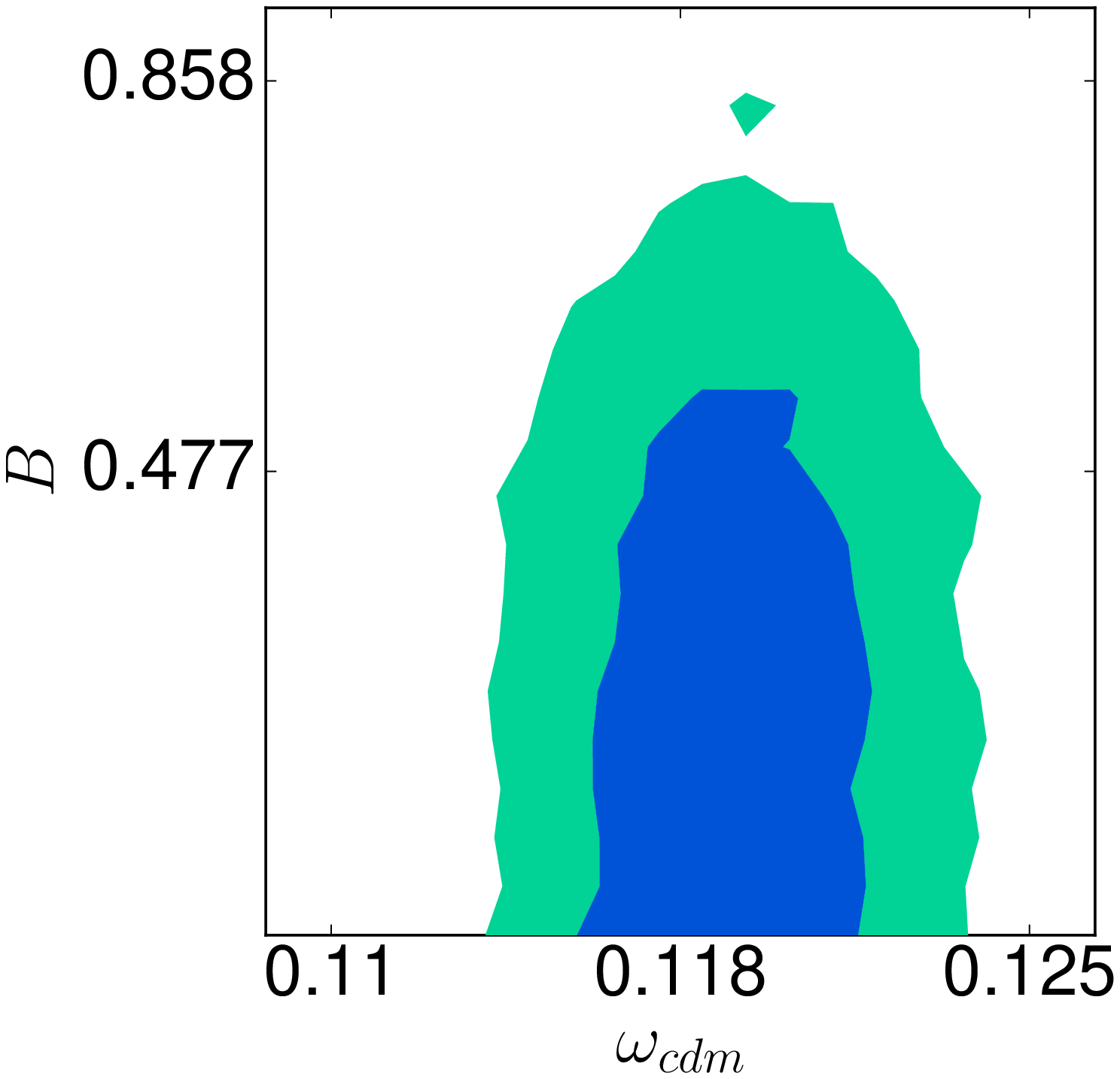}
\epsfxsize=1.15in\epsfbox{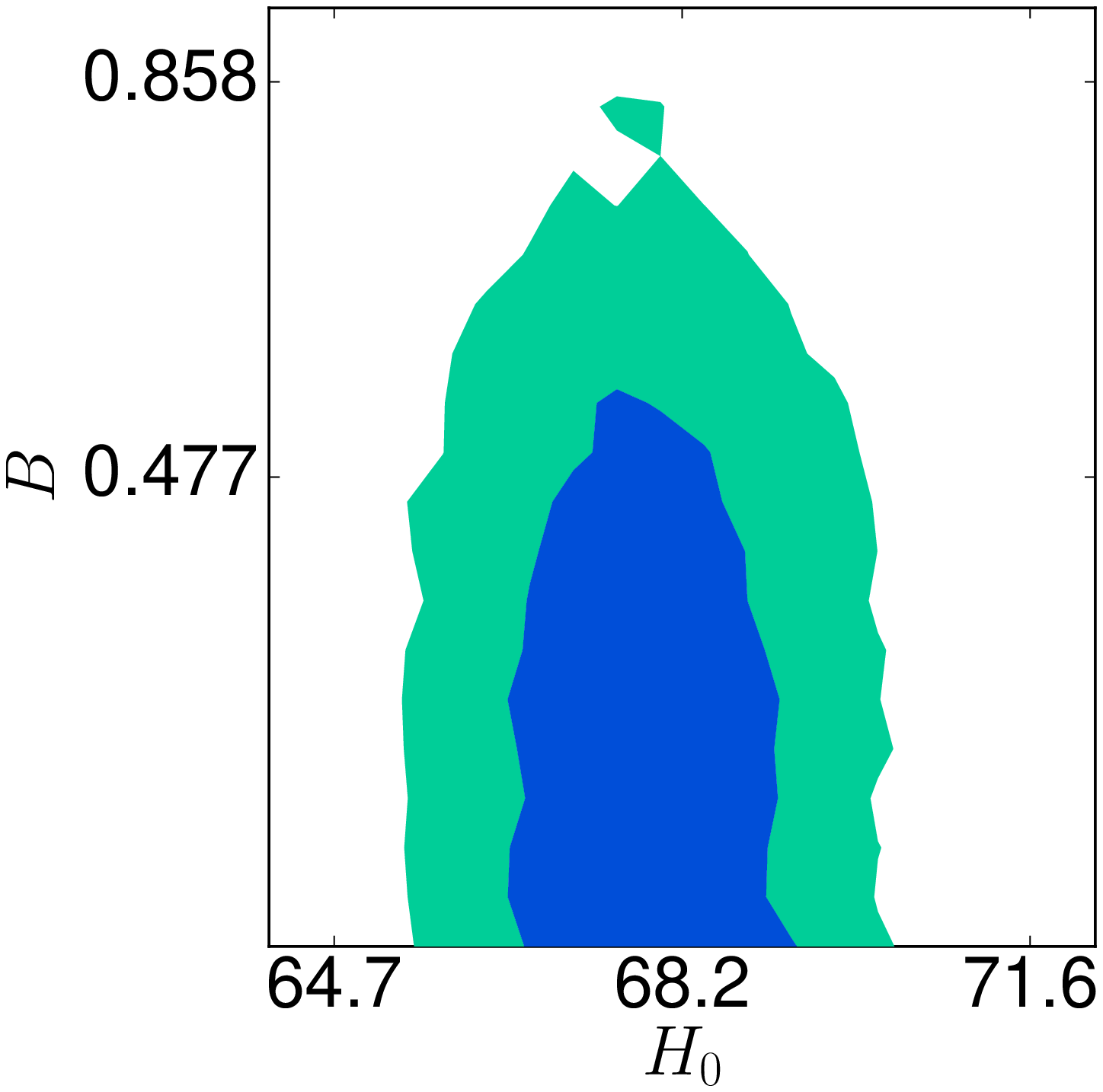}
\epsfxsize=1.15in\epsfbox{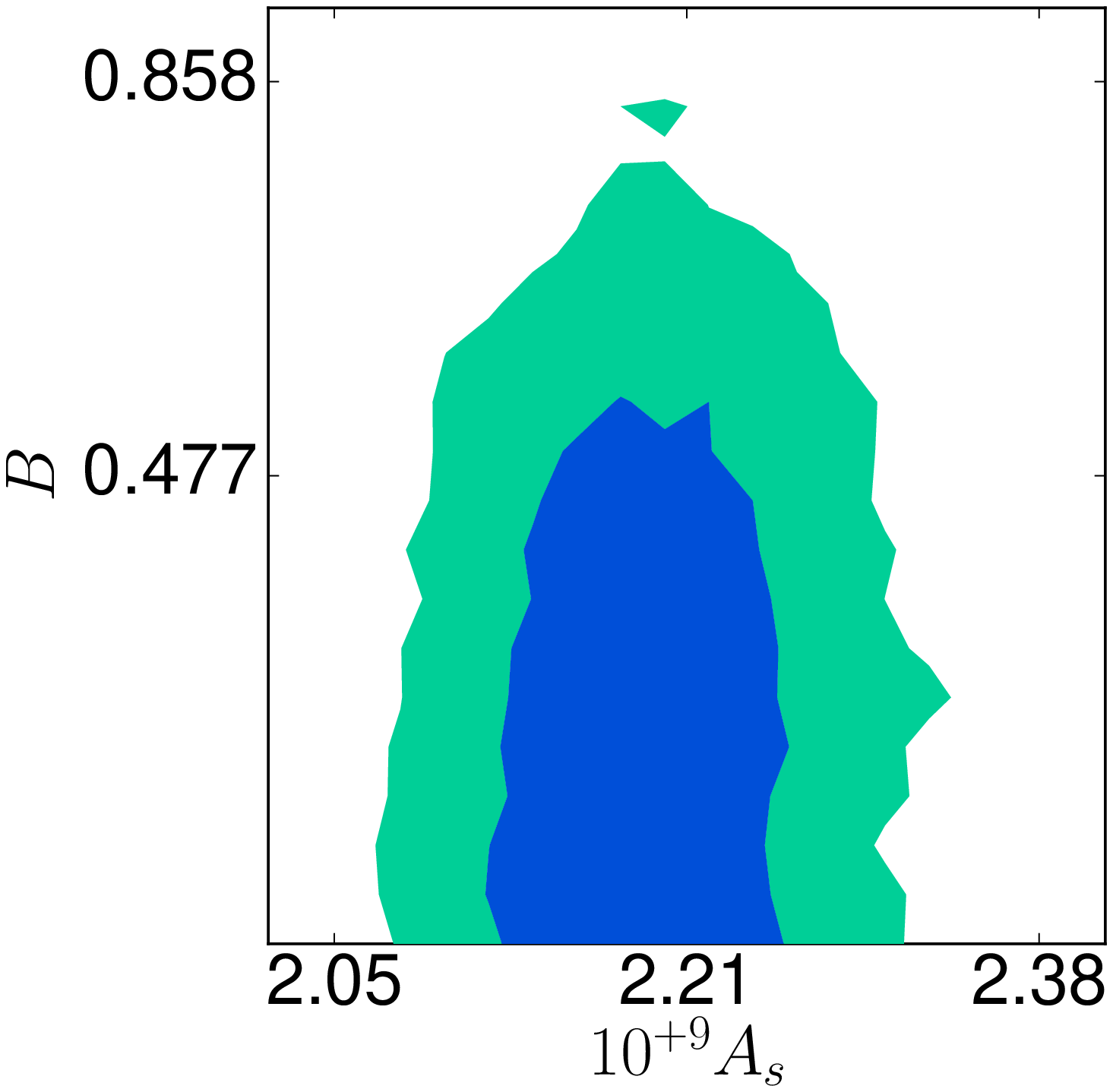}
\epsfxsize=1.15in\epsfbox{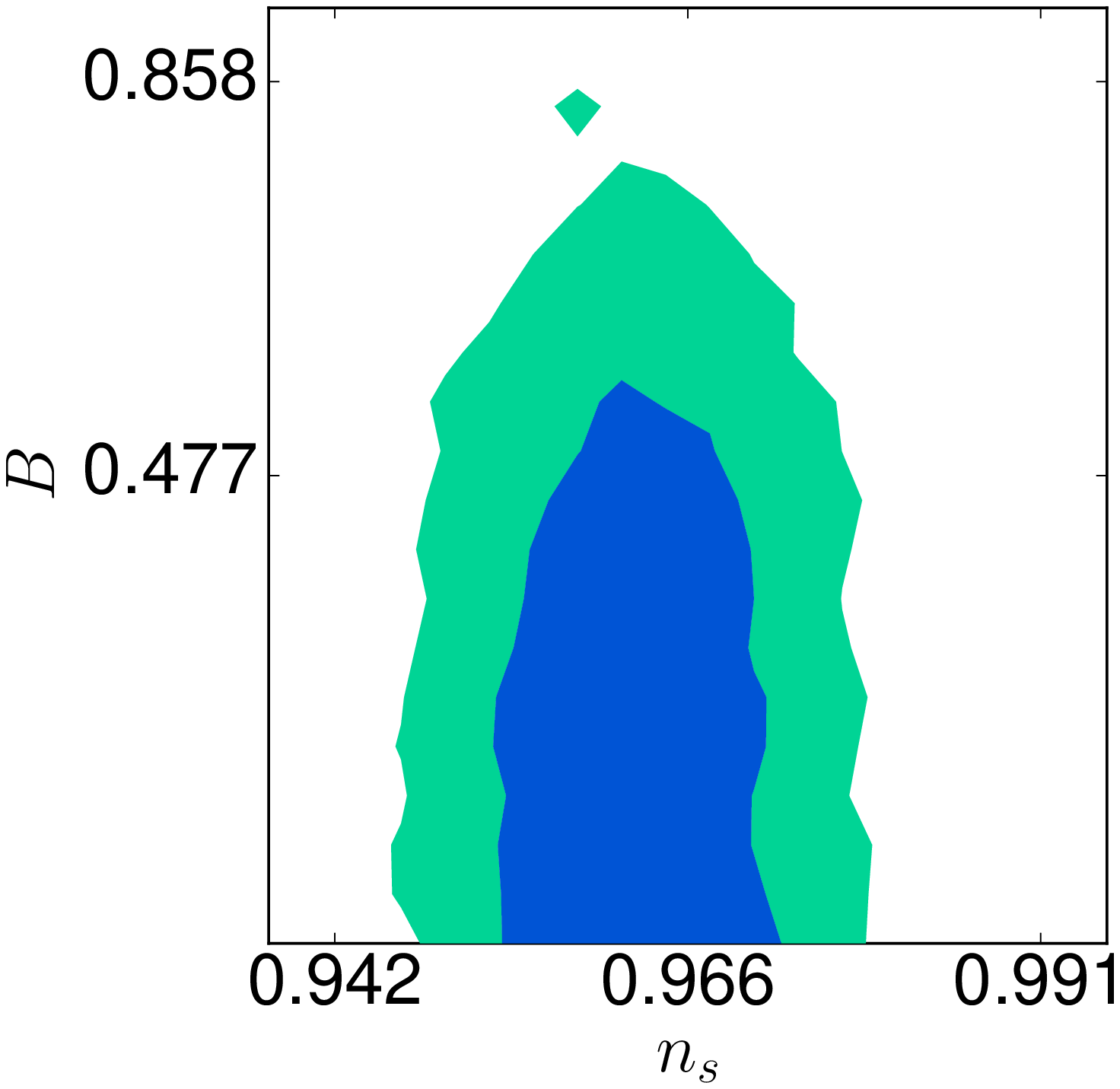}
\epsfxsize=1.15in\epsfbox{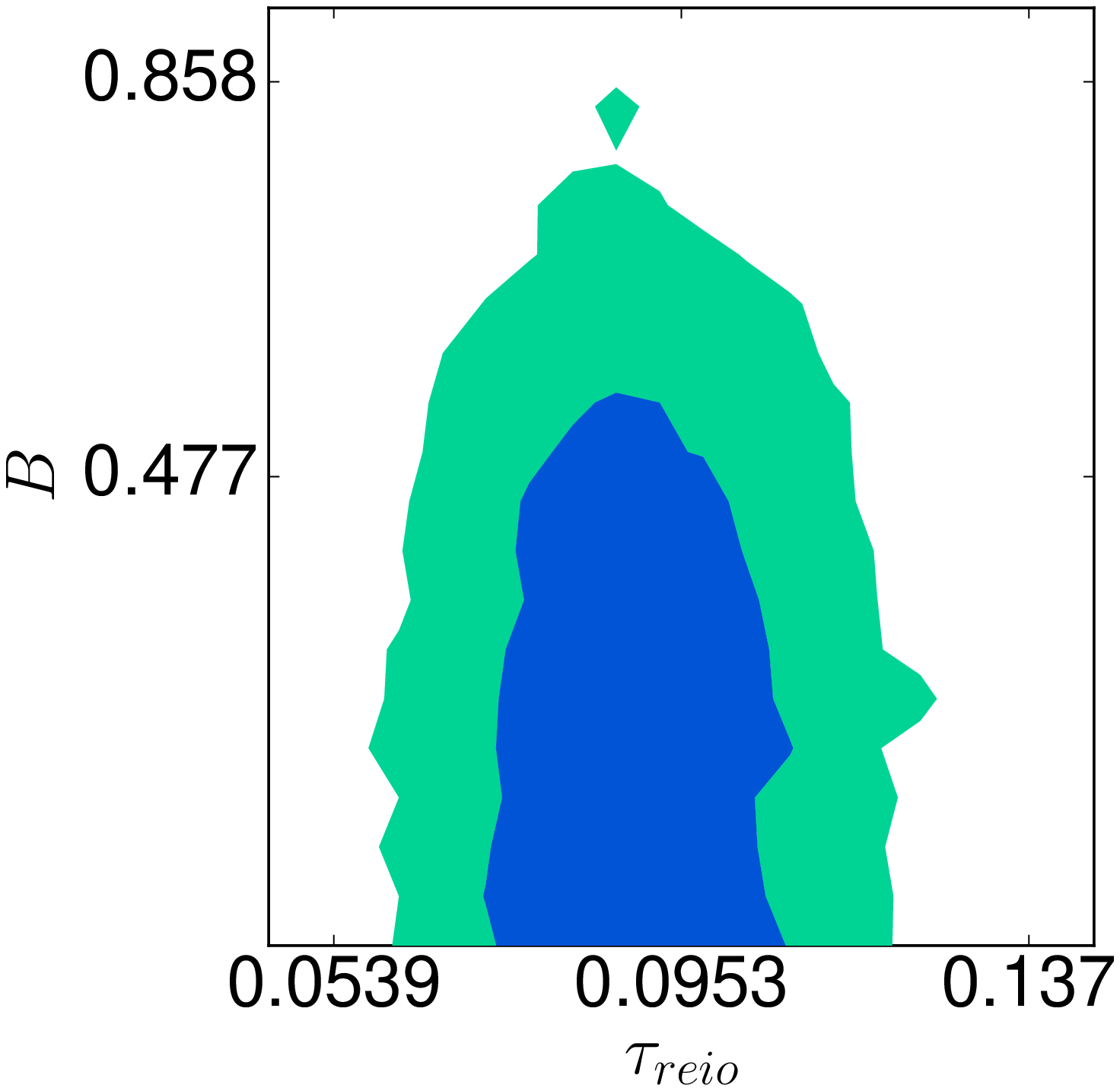}
}
\centerline{\epsfxsize=1.15in\epsfbox{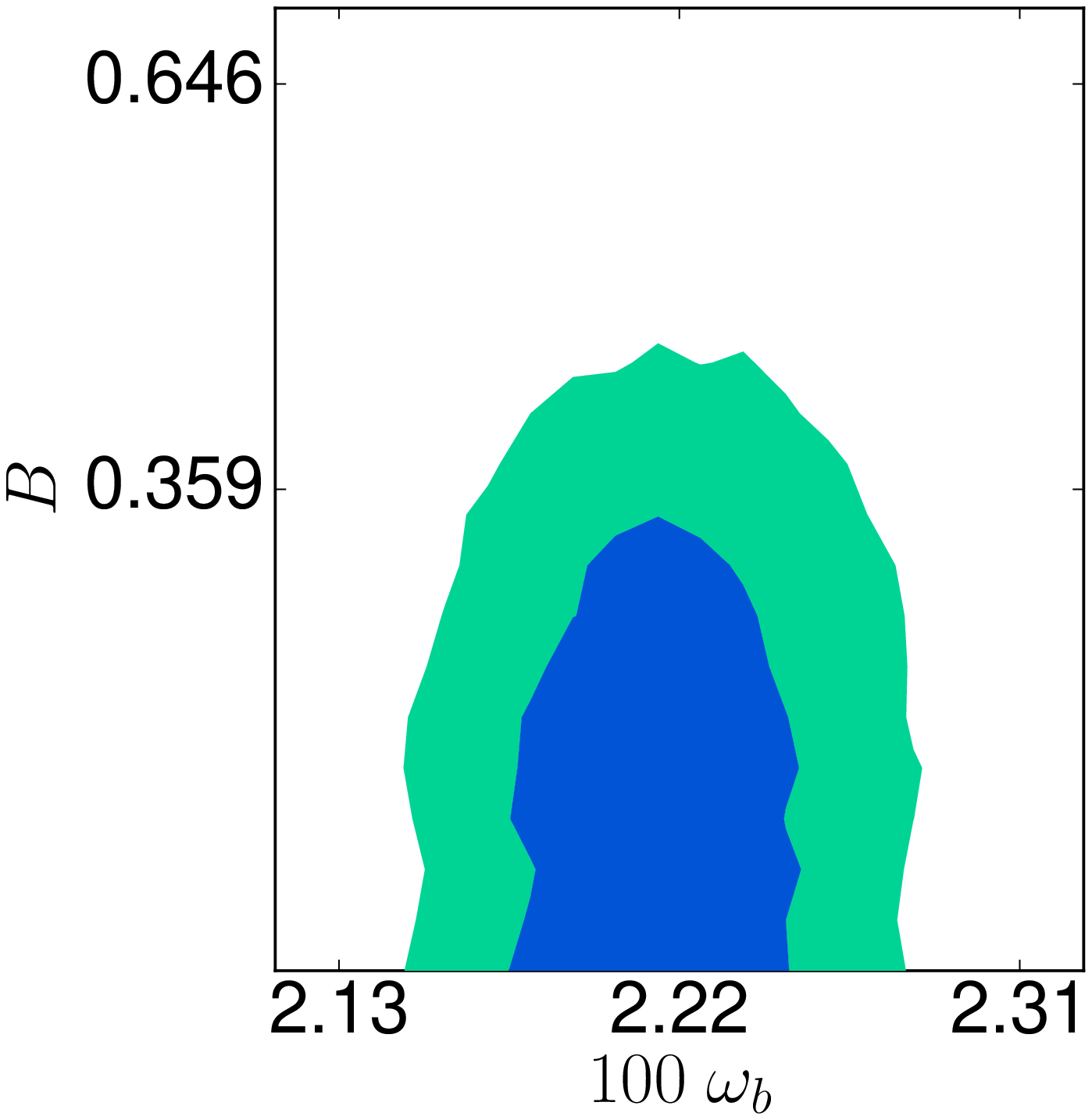}
\epsfxsize=1.15in\epsfbox{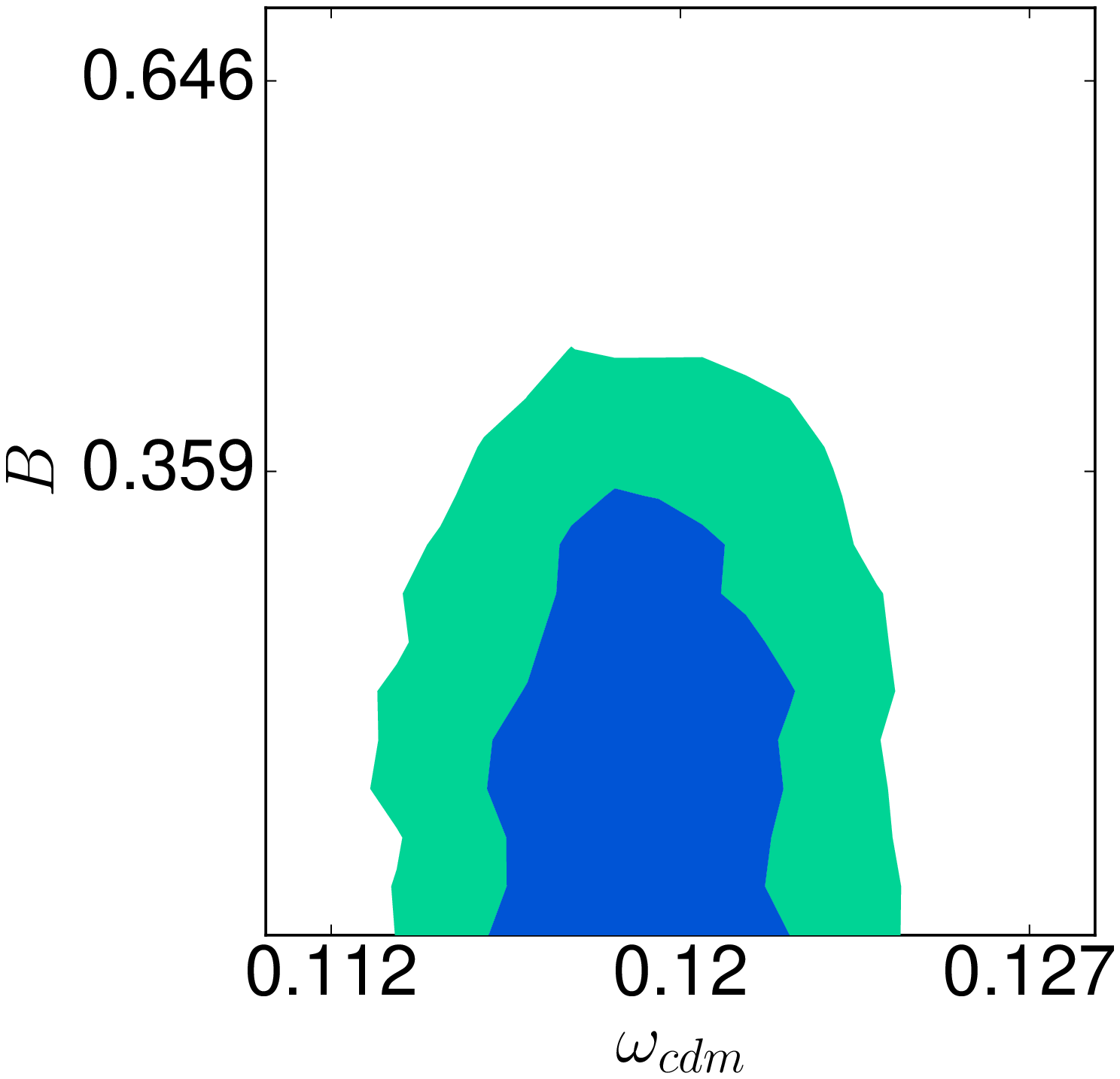}
\epsfxsize=1.15in\epsfbox{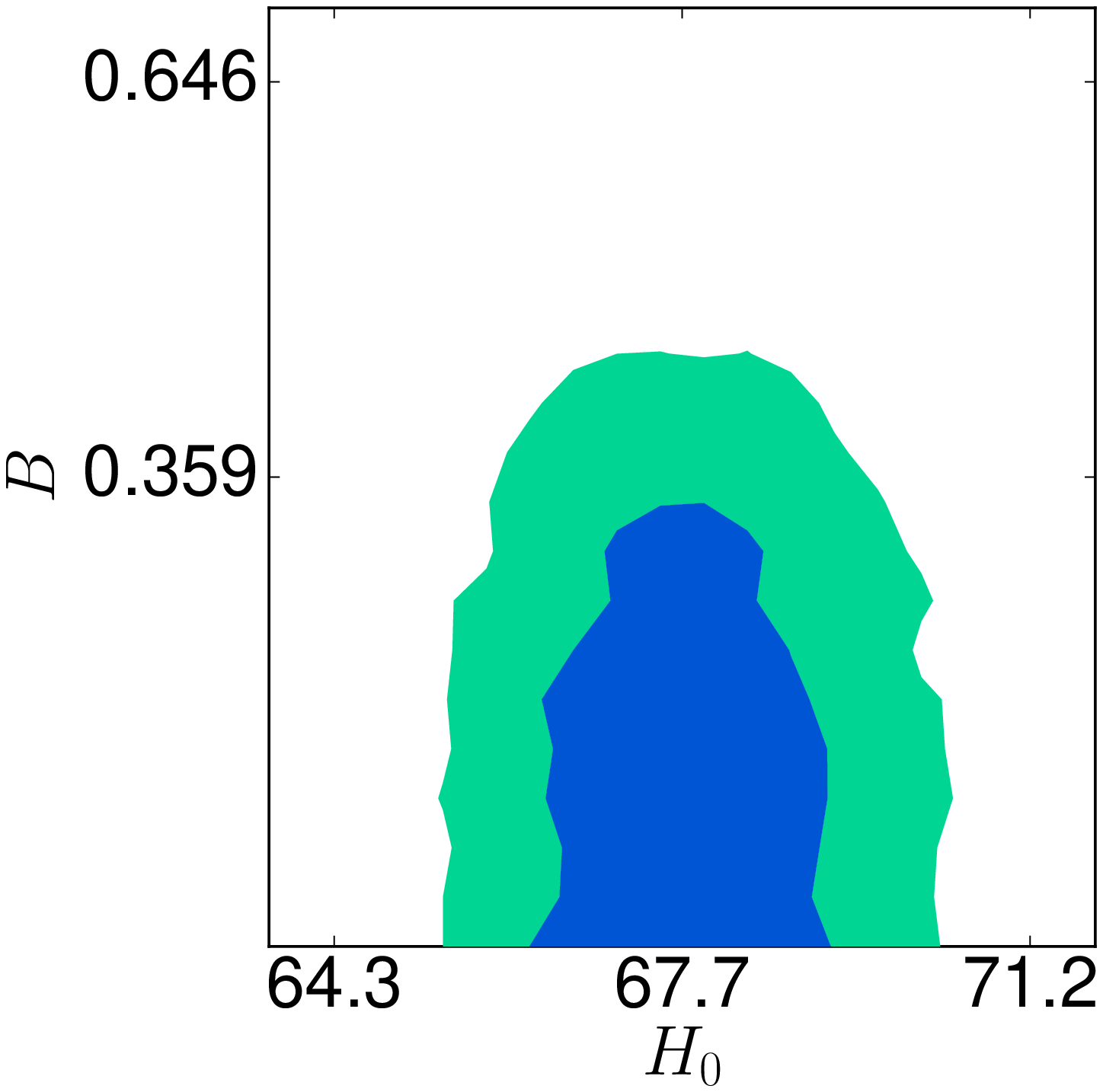}
\epsfxsize=1.15in\epsfbox{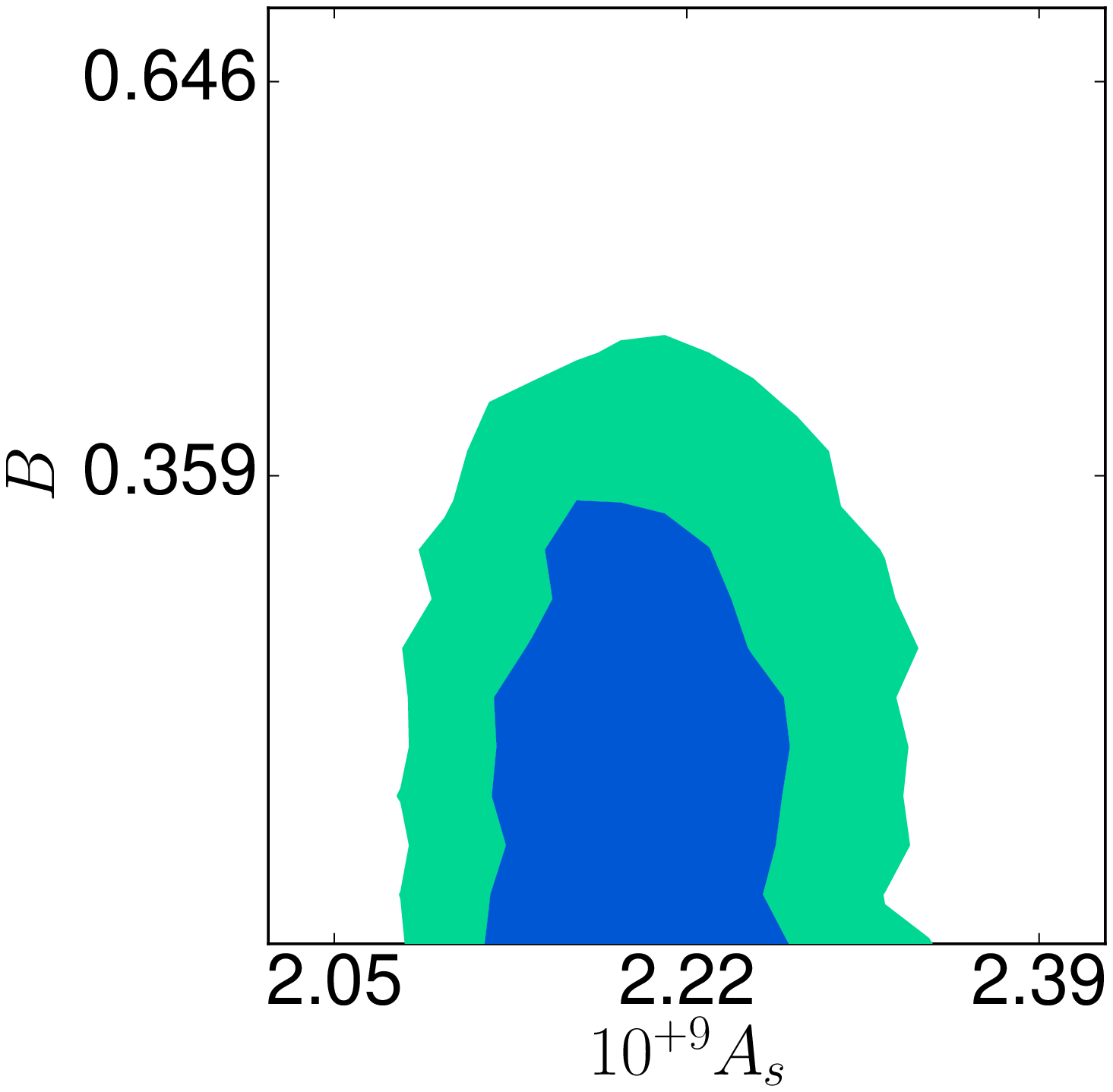}
\epsfxsize=1.15in\epsfbox{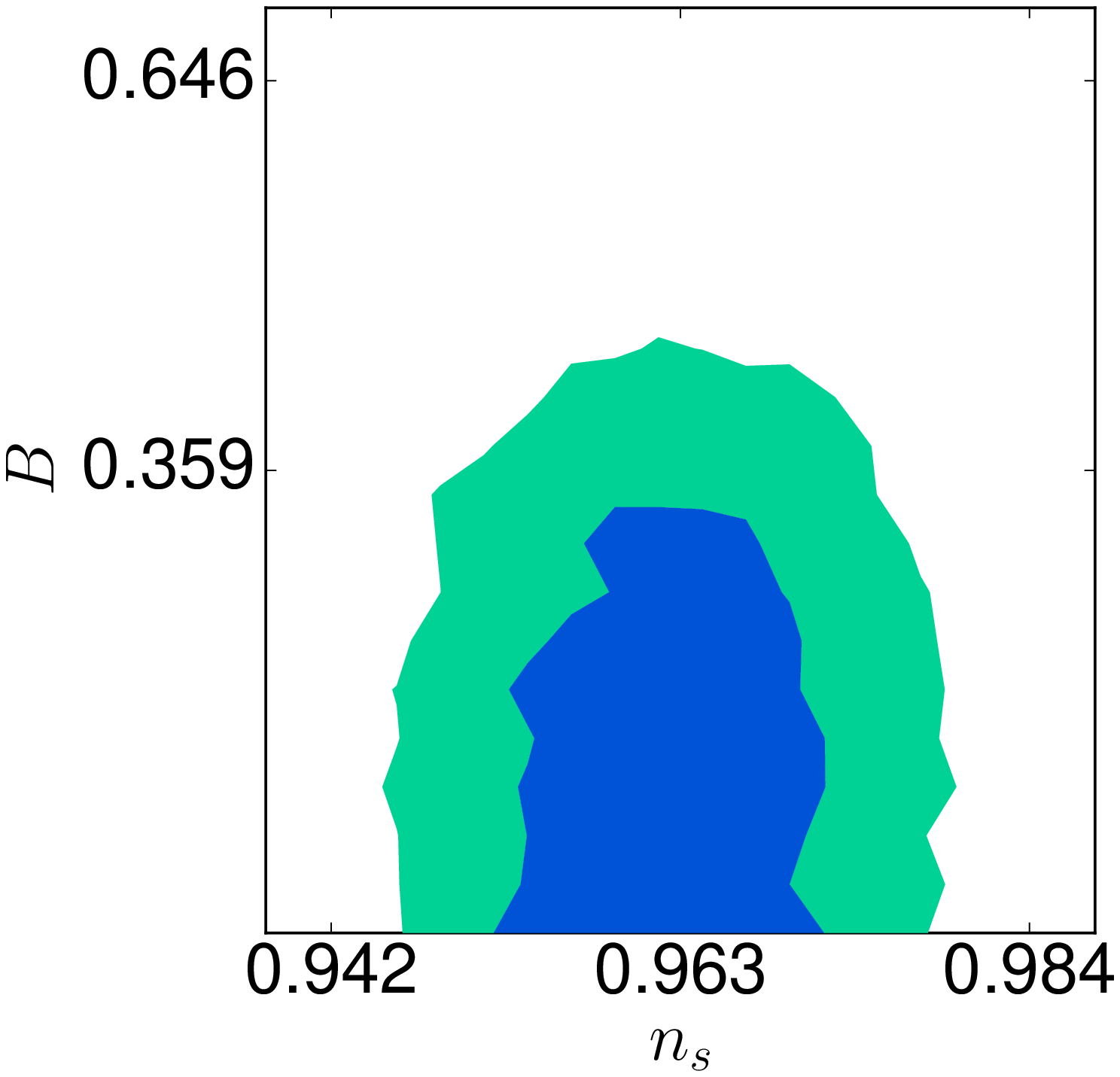}
\epsfxsize=1.15in\epsfbox{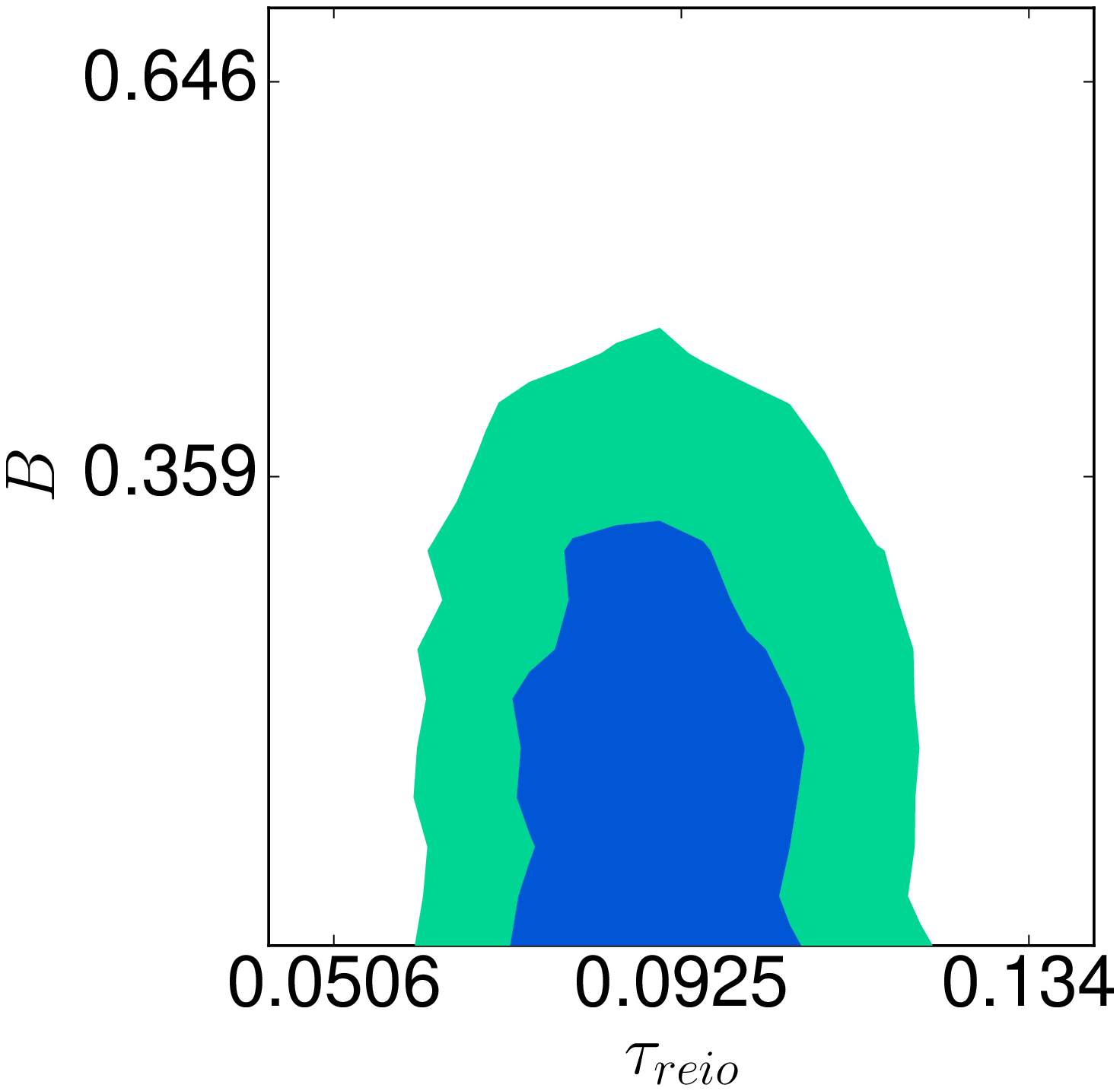}
}
\centerline{\epsfxsize=1.15in\epsfbox{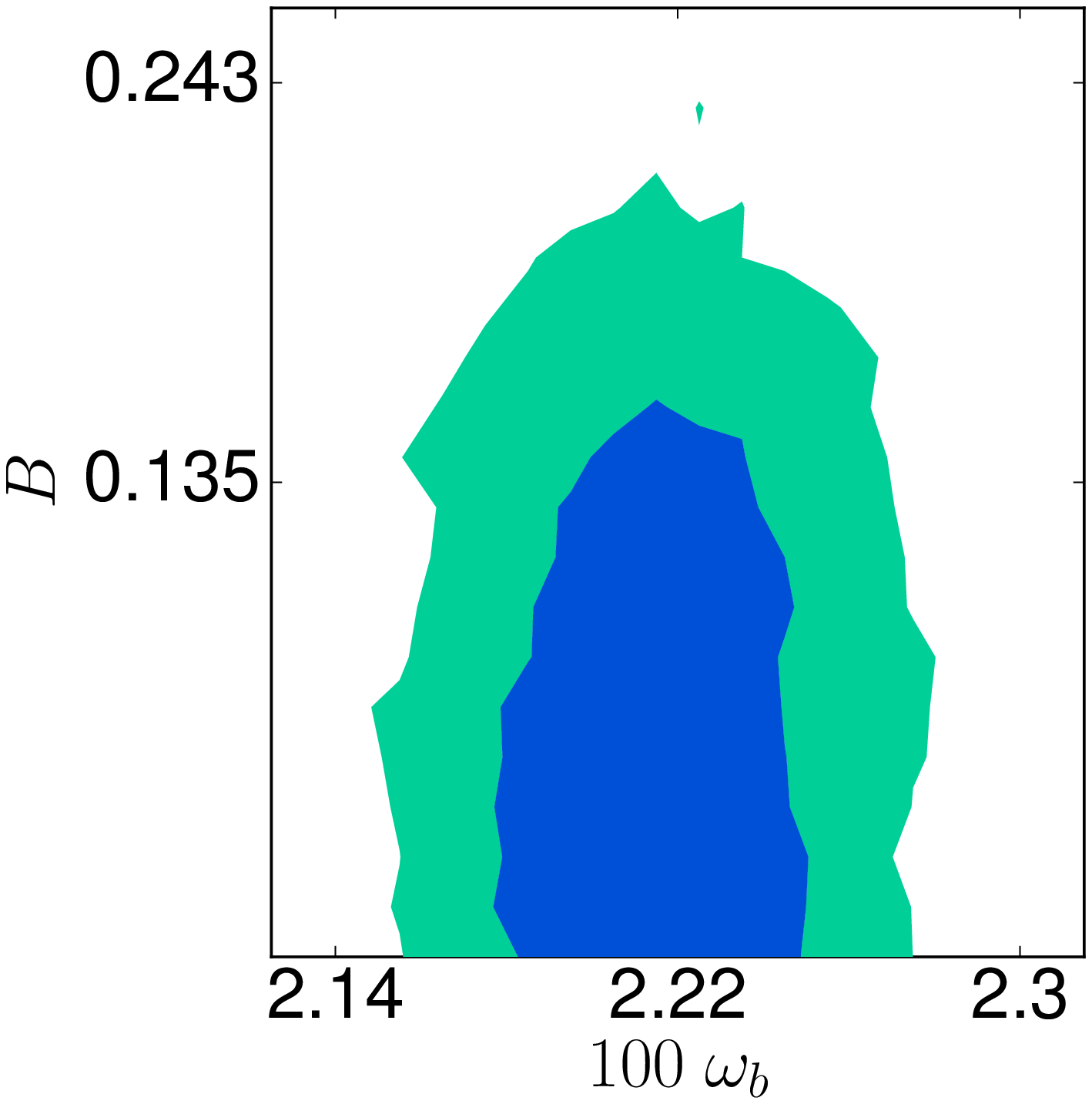}
\epsfxsize=1.15in\epsfbox{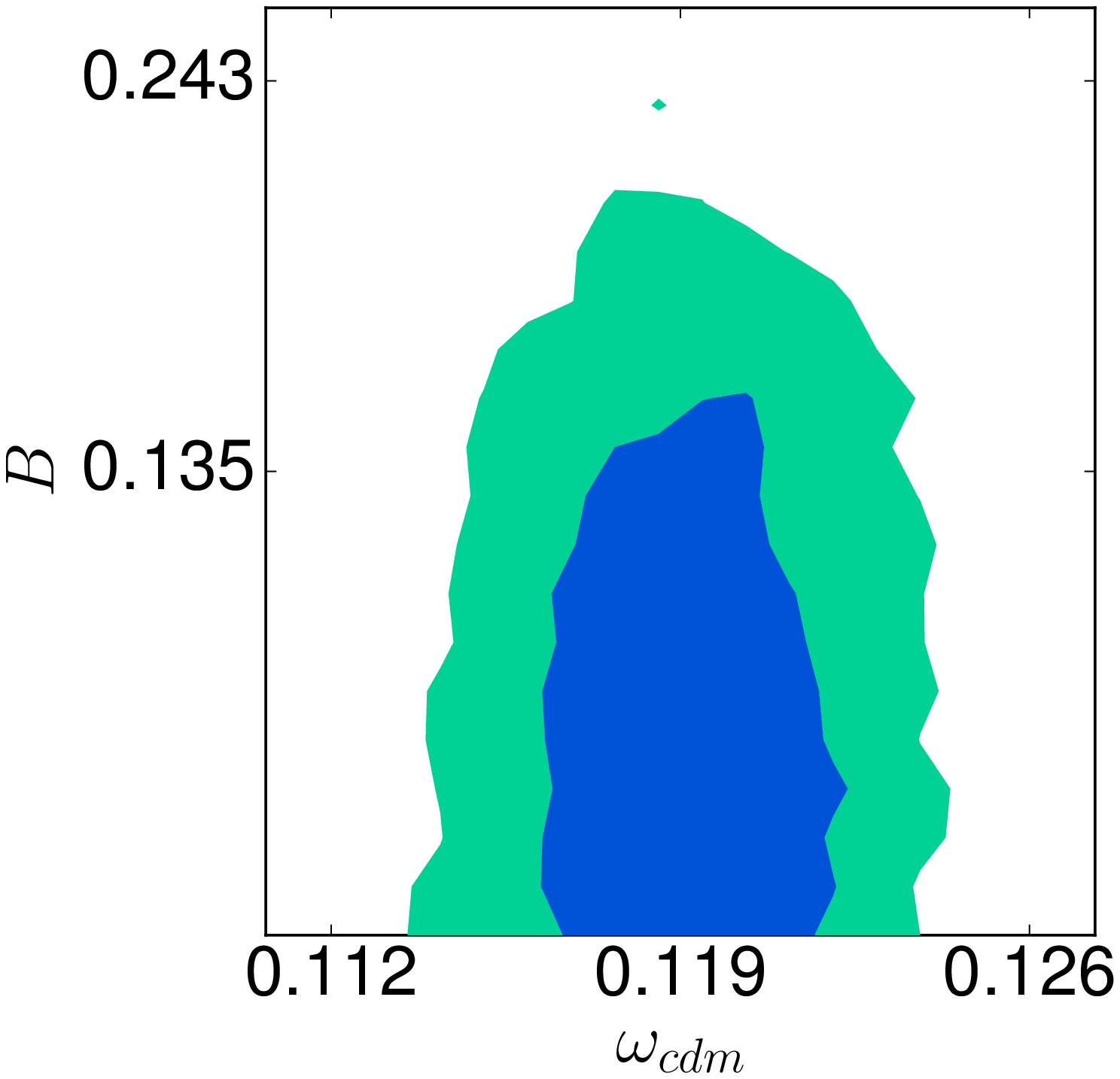}
\epsfxsize=1.15in\epsfbox{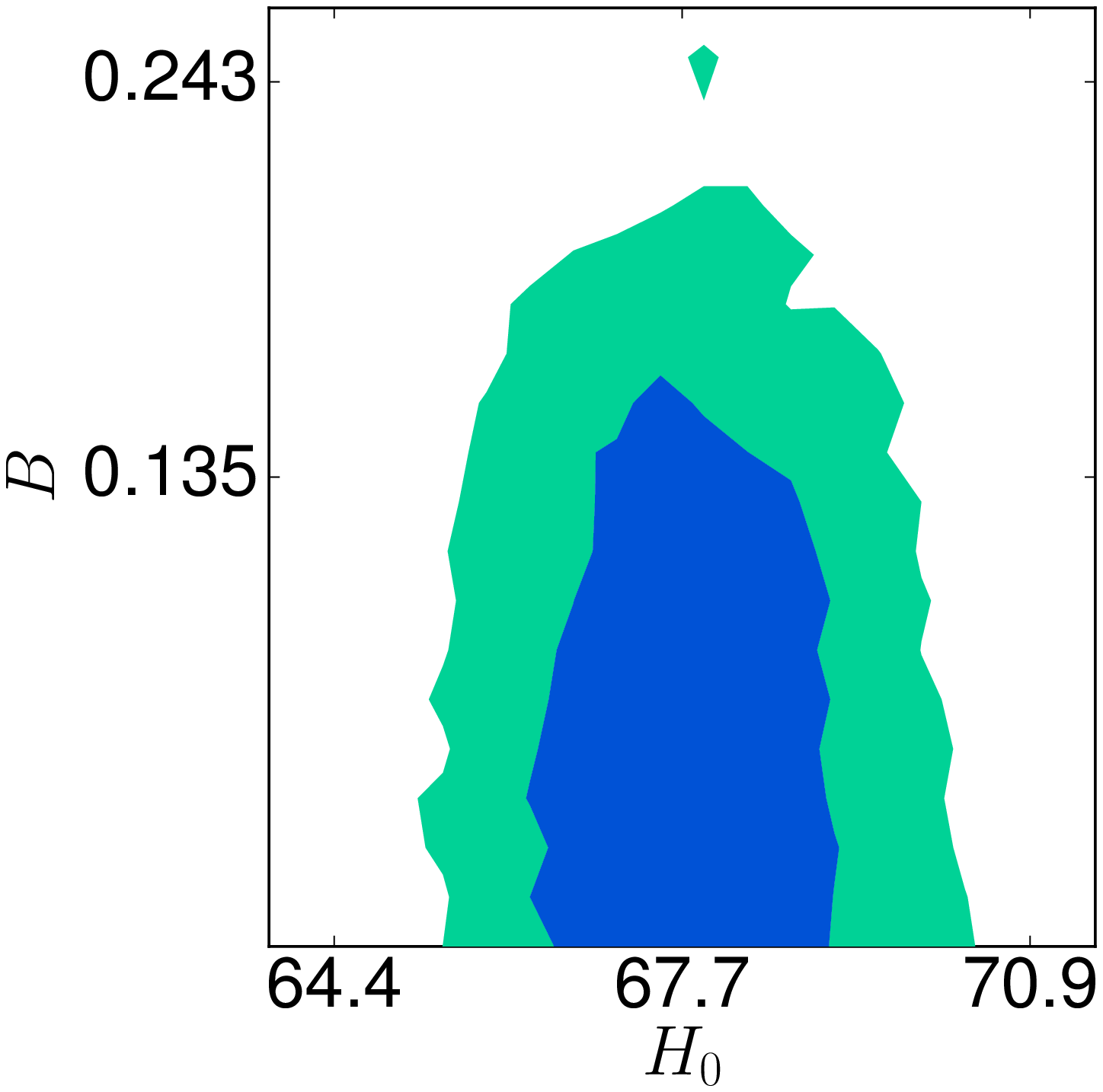}
\epsfxsize=1.15in\epsfbox{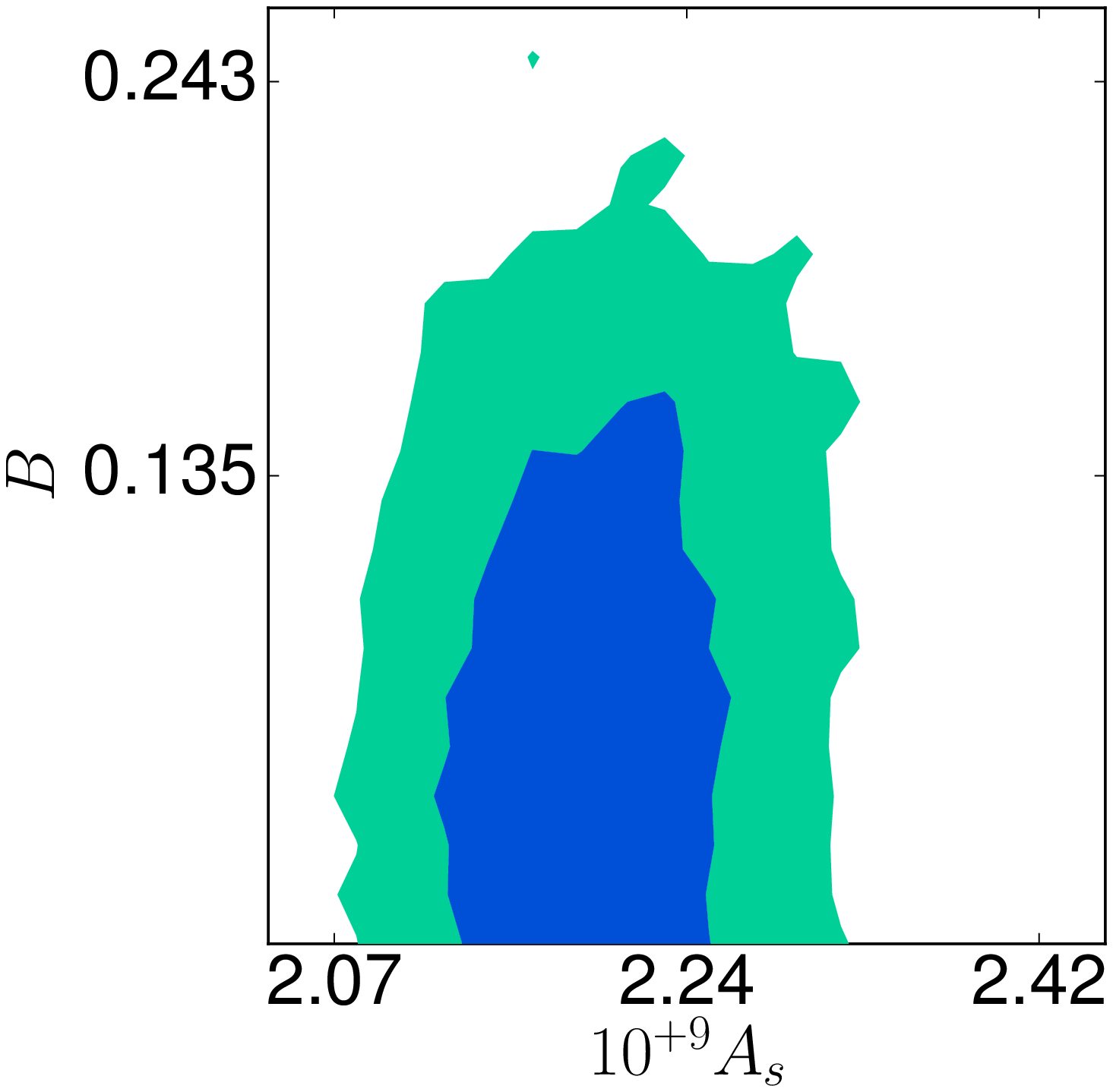}
\epsfxsize=1.15in\epsfbox{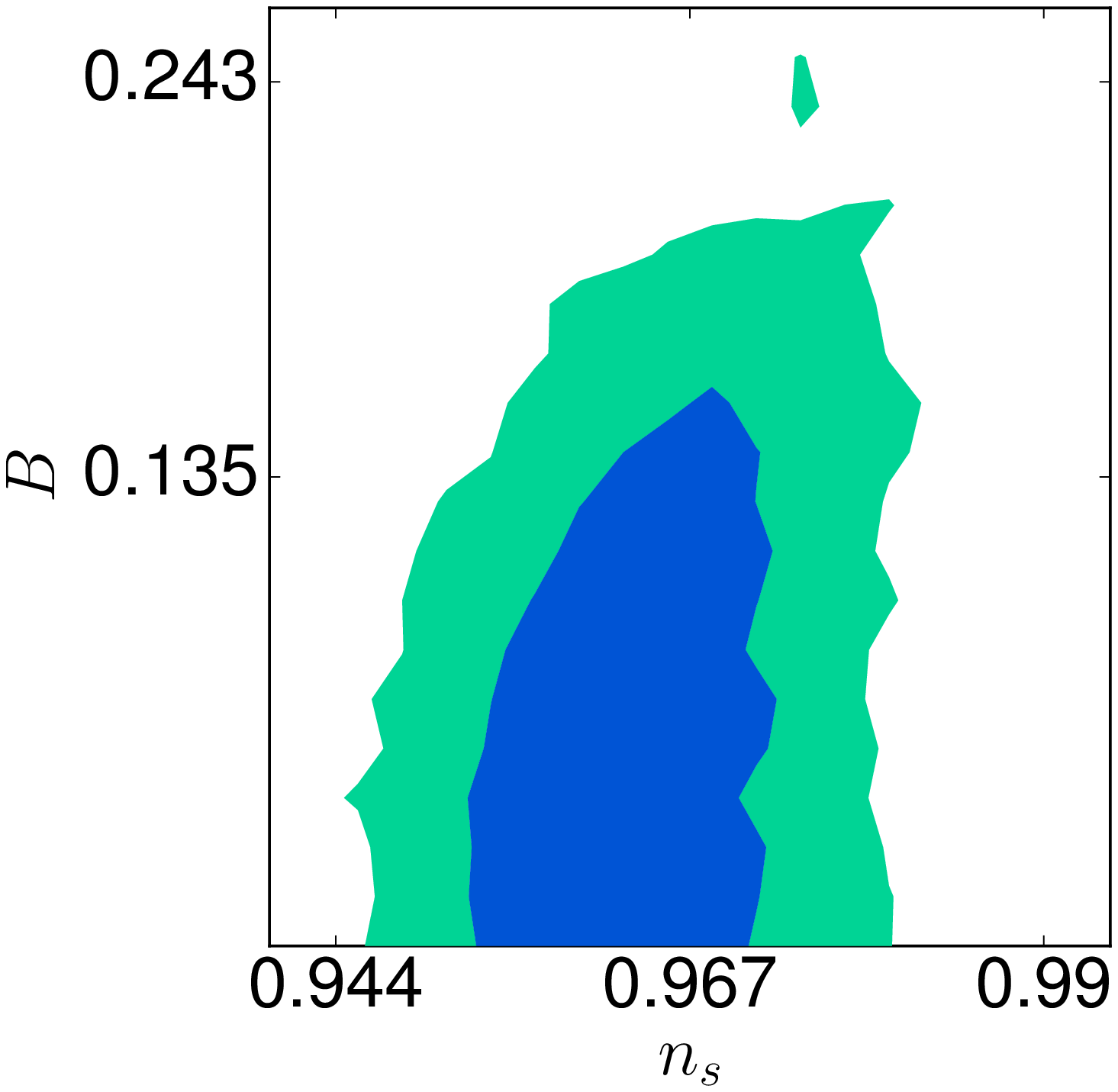}
\epsfxsize=1.15in\epsfbox{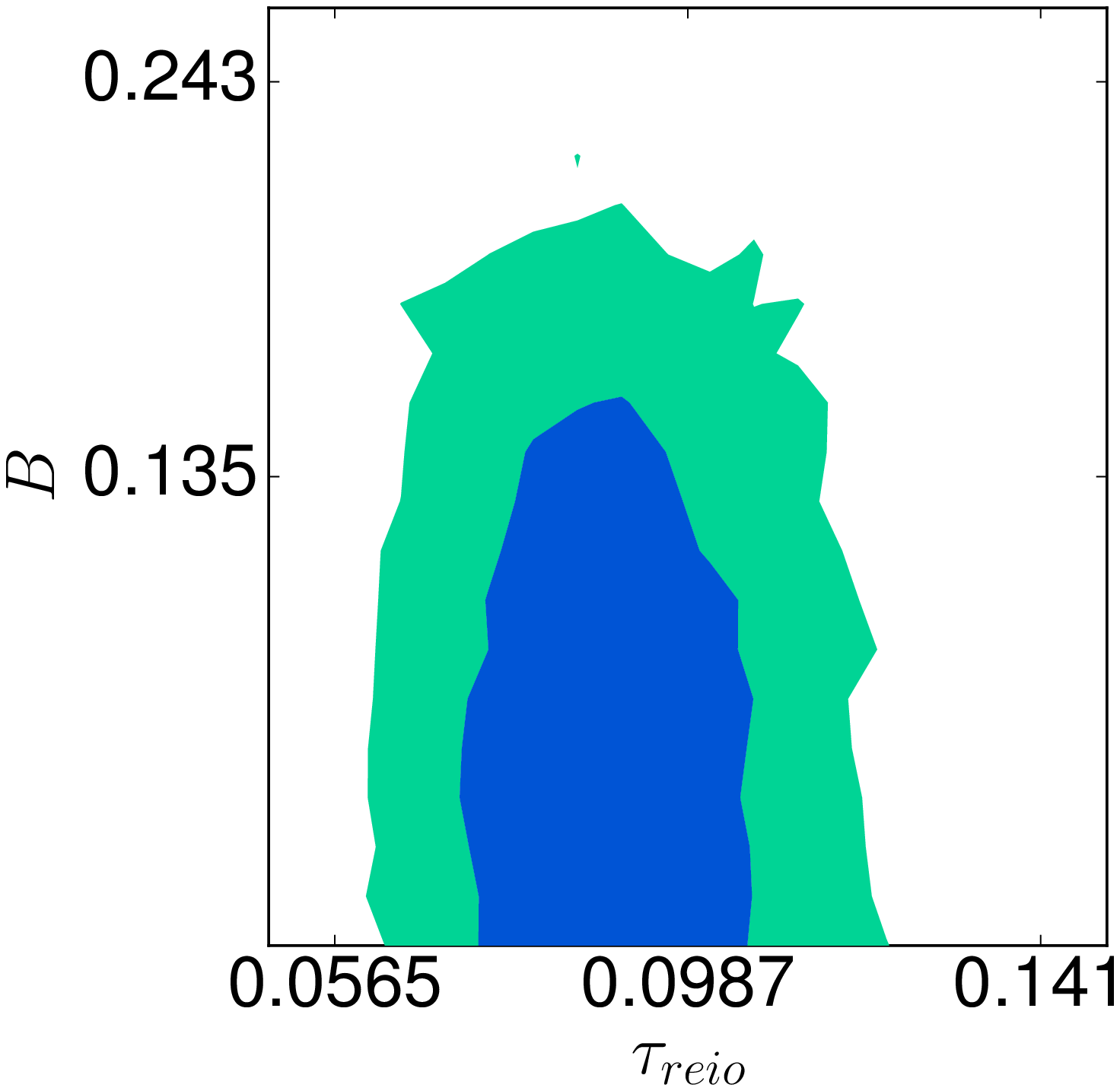}
}
\caption{68\% and 95\% confidence regions of the field strength, $B_0$
 (in units of nG), versus the cosmological parameters of the 
 $\Lambda$CDM model. The  magnetic
spectral indices are  $n_B=-2.9$ ({\it upper panel}), $n_B=-2.5$ ({\it middle panel}), and $n_B=-1.5$ ({\it lower panel}).}
\label{fig5}
\end{figure}

In figure \ref{fig5} the two-dimensional marginalized posterior
probability density distributions ($B_0$ versus the cosmological parameters)
are shown, for $n_B=-2.9$, $n_B=-2.5$, and $n_B=-1.5$. 
Most of the cosmological parameters are determined
independently of $B_0$; however, the scalar spectral tilt, $n_s$, is
weakly correlated with $B_0$ because $B_0$ suppresses the power spectrum
at high multipoles by $e^{-2\Delta\tau}$, which can be partly
compensated by increasing $n_s$.

\section{Conclusions}
\setcounter{equation}{0}
\label{sec_conc}

Dissipation of the magnetic fields in the post-decoupling
era heats the IGM and delays recombination of hydrogen atoms. This effect can be detected as the
extra optical depth to scattering of CMB photons. 
A qualitatively new
effect is that it affects only the total optical depth to decoupling, 
which is determined by the temperature data together with the lensing data, but not
the optical depth to reionization 
inferred from the low-$\ell$ polarization data ($\ell\lesssim 10$).
 Using the  2013 Planck data including the CMB lensing data, together
 with the
 likelihood of the low-$\ell$ polarization data from  WMAP as derived by the Planck collaboration,  we find
 no evidence for the effect of dissipation of (non-helical) magnetic
 fields in the CMB data. The  95\%~CL upper bounds are $B_0<0.63$, 0.39,
 and 0.18~nG for $n_B=-2.9$, $-2.5$, and $-1.5$, respectively. These
 limits are stronger than the previous limits
 that did not use the effect of fields on ionization history of the universe.

Inverse Compton scattering of CMB photons by hot electrons in the IGM
heated by dissipating magnetic fields in the post decoupling era leads
to a $y$-type spectral distortion \cite{jko2}. For negative
magnetic spectral indices, the Compton $y$ parameter is well
approximated by (see footnote 1 for the update from ref.~\cite{kuko})
\begin{eqnarray}
y(n_B, B_0)&=&1.2194\times 10^{-5}\left(\frac{B_0}{\rm nG}\right)^{1.7263}(-n_B)^{0.3602}
\nonumber\\
&&-1.2155\times 10^{-5}\left(\frac{B_0}{\rm nG}\right)^{1.7260}(-n_B)^{0.3619}{\rm e}^{9.3978\times 10^{-9}(-n_B)^{10.9842}}.
\end{eqnarray}
For the upper limits on the magnetic field amplitudes derived in this
paper, the resulting $y$-type distortions are 
$y<10^{-9}$, $4\times10^{-9}$, and $10^{-9}$
 for $n_B=-2.9$, $-2.5$, and $-1.5$, respectively. These limits are well
 below the contribution from the thermal Sunyaev-Zel'dovich effect in galaxy
clusters and groups, $y\approx 10^{-6}$ \cite{Refregier:2000xz}, and
thus are negligible.

We expect that this new method would yield much improved
limits, once the full Planck data sets including high-$\ell$
polarization are used, as they provide significantly better measurement
of the CMB lensing (which fixes $A_s$ thus breaking degeneracy between
$A_s$ and $e^{-2\Delta\tau}$), and potentially measure the extra
polarization at $100\lesssim \ell\lesssim 400$ generated by dissipation
of decaying MHD turbulence.

\section{Acknowledgements}
We would like to thank J. Chluba for very useful discussions.
KEK  would like to thank the Max-Planck-Institute for Astrophysics and the Perimeter Institute for Theoretical Physics for
hospitality where  part of this work was done. This research was supported in part by Perimeter Institute for Theoretical Physics. 
Research at Perimeter Institute is supported by the Government of Canada through Industry Canada and by the Province of Ontario through the Ministry of Economic Development \& Innovation. KEK acknowledges financial
support by Spanish Science Ministry grants FIS2012-30926 and
CSD2007-00042. 
We acknowledge the use of the Legacy Archive for
Microwave Background Data Analysis (LAMBDA). Support for LAMBDA is
provided by the NASA Office of Space Science.


\bibliography{references}
\end{document}